\documentclass{emulateapj} \usepackage{apjfonts}
\newcommand\xone{x_1}
\newcommand\xtwo{x_2}
\newcommand\cs{c_s}
\newcommand\Pbar{\Phi_{\rm bar}}

\newcommand\Msun{\; {\rm M}_{\odot}}
\newcommand\kms{\; {\rm km}\;{\rm s}^{-1}}
\newcommand\kpc{\;{\rm kpc}}
\newcommand\freq{\kms\kpc^{-1}}

\newcommand\yr{\; {\rm yr}}

\newcommand\pc{\;{\rm pc}}

\newcommand\Surf{\Msun\;{\rm pc^{-2}}}
\newcommand\dunit{\Msun\;{\rm kpc^{-3}}}
\newcommand\MBH{M_{\rm BH}}
\newcommand\Macc{M_{\rm acc}}
\newcommand\Rmax{R_{\rm max}}
\newcommand\Rmin{R_{\rm min}}
\newcommand\RCR{R_{\rm CR}}
\newcommand\ILR{R_{\rm ILR}}
\newcommand\IILR{R_{\rm IILR}}
\newcommand\OILR{R_{\rm OILR}}
\newcommand\Omb{\Omega_{b}}
\newcommand\rhobar{\rho_{\rm bar}}
\newcommand\rhobul{\rho_{\rm bul}}
\newcommand\vR{v_R}
\newcommand\vphi{v_\phi}
\newcommand\ip{i_p}
\newcommand\ish{i_{\rm sh}}
\newcommand\vperp{v_\bot}
\newcommand\vpara{v_\|}
\newcommand\Mperp{{\mathcal M}_\bot}
\newcommand\Mpara{{\mathcal M}_\|}
\newcommand\aSig{\langle\Sigma\rangle}
\newcommand\bSig{\Sigma_{\rm avg}}
\newcommand\pSig{\Sigma_{\rm peak}}
\newcommand\Rring{R_{\rm ring}}
\newcommand\Sring{\Sigma_{\rm ring}}
\newcommand\Mdot{\dot{M}}
\newcommand\fx{f_x}
\newcommand\fy{f_y}
\newcommand\fr{f_R}
\newcommand\fphi{f_\phi}
\newcommand\Aunit{\Msun\yr^{-1}}
\newcommand\simgt{\lower.5ex\hbox{$\; \buildrel > \over \sim \;$}}
\newcommand\simlt{\lower.5ex\hbox{$\; \buildrel < \over \sim \;$}}

\shorttitle{Nuclear Structures in Barred Galaxies}
\shortauthors{Kim et al.}

\begin{document}

\title{Central Regions of Barred Galaxies: Two-Dimensional
Non-self-gravitating Hydrodynamic Simulations}
\author{Woong-Tae Kim\altaffilmark{1,2,3}, Woo-Young Seo\altaffilmark{1,2},
James M.\ Stone\altaffilmark{4},
Doosoo Yoon\altaffilmark{2,5},
and Peter J.\ Teuben\altaffilmark{6}}
\affil{$^1$Center for the Exploration of the Origin of the Universe
(CEOU), Astronomy Program, Department of Physics \& Astronomy, Seoul
National University, Seoul 151-742, Republic of Korea}
\affil{$^2$FPRD, Department of Physics \& Astronomy, Seoul
National University, Seoul 151-742, Republic of Korea}
\affil{$^3$Institute for Advanced Study, Einstein Drive, Princeton,
NJ 08540, USA}
\affil{$^4$Department of Astrophysical Sciences, Princeton University,
Princeton, NJ 08544, USA}
\affil{$^5$Department of Astronomy,
University of Wisconsin-Madison, Madison, WI 53706, USA}
\affil{$^6$Department of Astronomy, University of Maryland, College
Park, MD 20742, USA}
\email{wkim@astro.snu.ac.kr}
\slugcomment{Accepted for Publication in the ApJ}

\begin{abstract}
The inner regions of barred galaxies contain substructures such as off-axis
shocks, nuclear rings, and nuclear spirals.  These substructure may affect
star formation, and control the activity of a central black hole (BH)
by determining the mass inflow rate.
We investigate the formation and properties of such substructures
using high-resolution, grid-based hydrodynamic simulations.
The gaseous medium is assumed to be infinitesimally-thin, isothermal,
and non-self-gravitating.  The stars and dark matter are represented by
a static gravitational potential with four components: a stellar disk, the
bulge, a central BH, and the bar.
To investigate various galactic environments, we vary the gas sound speed
$\cs$ as well as the mass of the central BH $\MBH$.
Once the flow has reached a quasi-steady
state, off-axis shocks tend to move closer to the bar major axis as
$\cs$ increases. Nuclear rings shrink in size with increasing $\cs$,
but are independent of $\MBH$, suggesting that ring position is
not determined by the Lindblad resonances. Rings in low-$\cs$ models
are narrow since they are occupied largely by gas on
$\xtwo$-orbits and well decoupled from nuclear spirals, while they
become broad because of large thermal perturbations in high-$\cs$
models.  Nuclear spirals persist only when either $\cs$ is small or
$\MBH$ is large; they would otherwise be destroyed completely by the
ring material on eccentric orbits.  The shape and strength of
nuclear spirals depend sensitively on $\cs$ and $\MBH$ such that
they are leading if both $\cs$ and $\MBH$ are small, weak trailing
if $\cs$ is small and $\MBH$ is large, and strong trailing if both
$\cs$ and $\MBH$ are large. While the mass inflow rate toward the
nucleus is quite small in low-$\cs$ models because of the
presence of a narrow nuclear ring, it becomes larger than
$0.01\Aunit$ when $\cs$ is large, providing a potential explanation of nuclear
activity in Seyfert galaxies.
\end{abstract}

\section{Introduction}\label{sec:intro}

Stellar bars play an important role in the dynamical evolution of
gas in galaxies.  By introducing a non-axisymmetric torque,
they produce interesting morphological substructures in the gaseous
medium, including a pair of dust lanes at the leading side of the
bar, a nuclear ring near the center, and nuclear spirals inside the
ring (e.g.,
\citealt{san76,rob79,sch81,van81,ath92b,pin95,but96,mar03a,mar03b,mar06}).
They also transport gas inward which can trigger starbursts in
the rings (e.g., \citealt{but86,gar91,hel94,bar95,mao01,maz08}) and
if the mass inflow extends all the way to the center, they may help
power active galactic nuclei
(AGN) (e.g., \citealt{shl90,reg99,kna00,lau04,van10}).

Since bar substructures represent a nonlinear response of the gas to
a non-axisymmetric gravitational potential, their formation and evolution is
best studied using direct numerical simulations.\footnote{\citet{eng00}
argued that physical properties of nuclear spirals
can be explained by the linear density-wave theories
(see also \citealt{mac04a}).}
There have been a
number of numerical studies on the gas dynamics in barred galaxies.
Based on the numerical scheme employed, they can be categorized
largely into two groups: (1) those using a smoothed particle
hydrodynamics (SPH) technique (e.g.,
\citealt{eng97,pat00,ann05,tha09}) and (2) those using a grid-based
algorithm (e.g., \citealt{ath92b,pin95,mac02,mac04b,reg03,reg04}).
The numerical results from these two approaches do not always agree
with each other, at least quantitatively, even if the model parameters
are almost identical. For instance, \citet{pin95} using the CMHOG
code on a cylindrical grid reported that the gas near the corotation
regions exhibits complex density features resulting from
Rayleigh-Taylor and/or Kelvin-Helmholtz instabilities, while these
structures are absent in the SPH simulations.
In addition, overall shapes and structures of dust lanes and nuclear rings
from CMHOG simulations are different from SPH results.

Some differences in the numerical results may be attributable to
relatively large numerical diffusion of a standard SPH method and
its inability to handle sharp discontinuities accurately
(e.g., \citealt{age07,pri08,rea10}).  However, after adopting and thoroughly
testing the CMHOG code as part of this work, we have found it contained a
serious bug in the way the gravitational forces due to the bar are added
to the hydrodynamical equations.  Thus, some of the discrepancies in the
flows computed by CMHOG and other codes are likely due to this bug.
We discuss this bug and its affect on the results reported in \citet{pin95}
in Section 2.2.

In this paper, we revisit the gas dynamics in barred galaxies using
a corrected version of the CMHOG code.  Our objectives are
three-fold. First, we wish to remedy the errors in \citet{pin95},
and to compute the formation of bar substructures with an accurate
shock-capturing grid code with the correct bar potential.  Second,
the morphology, shape, and strength of the bar substructures are
likely to depend on the gas sound speed and the shape of the
underlying gravitational potential (e.g., \citealt{eng00}).   Thus,
we report new models in which we include a central black hole (BH)
that greatly affects the gravitational potential in the central
regions, and we vary both the BH mass as well as the sound speed to
explore the dynamics in various galactic conditions.  Third, we
exploit advances in computational resources to compute models that
have more than an order of magnitude higher resolution than the
models in \citet{pin95}, with a grid resolution of $0.13$ pc in the
central regions.  This allows us to resolve details in the flow in
the nuclear regions, in particular the formation of nuclear rings
and nuclear spirals.

According to the most widely accepted theory, a nuclear ring forms
near the inner Lindblad resonance (ILR) when there is only one ILR,
as the gas outside (inside) ILR loses (gains) angular momentum and
accumulates there, while it forms in between the inner ILR and outer
ILR when there are two ILRs (e.g., \citealt{shl90,com96,but96}). On
the other hand, \citet{reg03} argued that the ring formation is more
deeply related to the existence of $\xtwo$-orbits rather than the
ILRs.  But, the arguments relying on either ILR or $\xtwo$-orbits do
not take into account the effect of thermal pressure.  Therefore, it is
important to explore to what extent the concepts of ILR or
$\xtwo$-orbits are valid in describing nuclear rings, especially
when the sound speed is large.

The formation, shape, and nature of nuclear spirals that may channel the
gas to the galaxy centers are also not well understood. Observations
using the \emph{Hubble Space Telescope} indicate that galaxies
having nuclear dust spirals are quite common (e.g.,
\citealt{mar03a,mar03b}). While most of such spirals are trailing, a
few galaxies including NGC 1241 and NGC 6902 reportedly possess
leading nuclear spirals \citep{dia03,gro03}. Although the linear
theory suggests that leading spirals are expected when there are two
ILRs (e.g., \citealt{mac04a}), they are absent in the numerical
models of \citet{pin95} computed with the CMHOG code, while the SPH models
of \citet{ann05} with self-gravity do form leading spirals.
The SPH models suffer from poor spatial resolution at the nuclear
regions as most particles gather around the rings.  By running
high-resolution simulations with a corrected version of CMHOG, we can
clarify the issues of the nuclear
spiral formation and related mass inflow rates to the galaxy center.

We in this work treat gaseous disks as being two-dimensional,
isothermal, non-self-gravitating, and unmagnetized, which introduces
a few caveats that need be noted from the outset.
By considering an infinitesimally-thin disk, we ignore gas motions
and associated dynamics along the direction perpendicular to the disk.
By imposing a point symmetry relative to the galaxy center, our
models do not allow for the existence of odd-$m$ modes, although this
appears reasonable since $m=2$ modes dominate in the problems
involving a galactic bar.
In addition, we are unable to capture the potential consequences of gaseous
self-gravity and magnetic stress that may not only cause
fragmentation of high-density nuclear rings but also affect mass inflow
rates to the galaxy center.
Nevertheless, these idealized models are
useful to isolate the effects of the gas sound speed and the mass of
a central BH on the formation of bar substructures and mass inflows.
Also, these models allow us to correct the results of previous CMHOG
calculations with incorrect bar forces.

This paper is organized as follows. In Section 2, we describe the
galaxy model, model parameters, and our numerical methods. In
Section 3, we present the results of simulations for off-axis shocks
and nuclear rings. The detailed properties of nuclear spirals are
presented in Section 4. In Section 5, we study the mass inflow rates
through the inner boundary obtained from our simulations. In Section
6, we conclude with a summary and discussion of our results and
their astronomical implications.

\begin{deluxetable}{lcc}
\tabletypesize{\footnotesize}
\tablewidth{0pt}
\tablecaption{Model Parameters\label{tbl:model}}
\tablehead{
\colhead{$\;\;\;\;\;$Model$\;\;\;\;\;$} &
\colhead{$\;\;\;\;\;$$\cs ~ ({\rm km\; s^{-1}})$$\;\;\;\;\;$} &
\colhead{$\;\;\;\MBH(0) (\Msun)\;\;\;$}
}
\startdata
cs05bh0 & 5  & 0  \\
cs05bh0t\tablenotemark{a} & 5  & 0  \\
cs10bh0 & 10 & 0  \\
cs15bh0 & 15 & 0  \\
cs20bh0 & 20 & 0  \\
cs20bh0t\tablenotemark{a} & 20 & 0  \\
\hline
cs05bh7 & 5  & 4$\times10^7$ \\
cs10bh7 & 10 & 4$\times10^7$ \\
cs15bh7 & 15 & 4$\times10^7$ \\
cs20bh7 & 20 & 4$\times10^7$ \\
cs20bh7t\tablenotemark{a} & 20 & 4$\times10^7$ \\
\hline
cs05bh8 & 5  & 4$\times10^8$ \\
cs10bh8 & 10 & 4$\times10^8$ \\
cs15bh8 & 15 & 4$\times10^8$ \\
cs20bh8 & 20 & 4$\times10^8$
\tablenotetext{a}{
The BH mass is varied with time as
$\MBH(t) = \MBH(0) + \int_0^t \Mdot(t^\prime)dt^\prime$
assuming that all the inflowing mass is added to the central BH.}
\end{deluxetable}

\section{Models and Methods\label{sec:model}}

We consider a uniform, isothermal, infinitesimally-thin, and
non-self-gravitating gas disk orbiting in a gravitational potential
$\Phi_{\rm ext}$ arising from various components of a barred galaxy.
The bar is assumed to rotate about the galaxy center with a fixed
pattern speed $\mathbf{\Omb}=\Omb\mathbf{\hat z}$. Therefore, it is
advantageous to solve the dynamical equations in cylindrical polar
coordinates ($R$, $\phi$) corotating with the bar in the $z=0$
plane. The equations of ideal hydrodynamics in this rotating frame
are
\begin{equation}\label{eq:con}
\left(\frac{\partial}{\partial t}  + \mathbf{u}\cdot\nabla \right) \Sigma
= - \Sigma\nabla\cdot\mathbf{u},
\end{equation}
\begin{equation}\label{eq:mom}
\left(\frac{\partial}{\partial t}  + \mathbf{u}\cdot\nabla \right) \mathbf{u}
= -\cs^2 \frac{\nabla \Sigma}{\Sigma} - \nabla \Phi_{\rm ext} + \Omb^2
\mathbf{R} - 2\mathbf{\Omb}\times \mathbf{u},
\end{equation}
where $\Sigma$, $\mathbf{u}$, and $\cs$ denote the surface density,
velocity in the rotating frame, and the sound speed in the gas,
respectively.
The third and fourth terms in the right hand side of equation (\ref{eq:mom})
represent the Coriolis and centrifugal forces, respectively, arising
from the coordinate transformation from the inertial to rotating frames.
The velocity $\mathbf{v}$ in the inertial frame is obtained from
$\mathbf{v} = \mathbf{u} + R\Omb\mathbf{\hat \phi}$.
In order to focus on the bar-driven gas dynamics, we do not consider
star formation and the associated gas recycling in the present work.

The external gravitational potential $\Phi_{\rm ext}$ consists of
four components: an axisymmetric stellar disk, a spherical bulge,
a non-axisymmetric bar, and a central supermassive BH.
Appendix \ref{sec:appen} describes the specific potential model
we employ for each component of the galaxy.
The bar pattern speed is taken to be $\Omb=33\freq$.
Without a central BH, our galaxy model is similar to those
in \citet{ath92a,ath92b} and \citet{pin95}.
The presence of a BH allows us to explore the effect of central
mass concentration on the formation of nuclear spirals
(e.g., \citealt{mac04b,tha09}).

\subsection{Models}

The real interstellar gas is multiphase and turbulent, with
temperatures differing by a few orders of magnitude
(e.g., \citealt{fie69,mck77,mck07}).  For simplicity,
we model this highly inhomogeneous gas using an isothermal equation of state
with an effective sound speed $\cs$ that includes a contribution
due to turbulent motions.  We have calculated 15 different models in which
we vary both $\cs$ and the initial mass $\MBH(0)$
of the central BH as parameters.
Table \ref{tbl:model} lists the properties of each calculation.
The sound speed is chosen to vary between 5 and $20\kms$.
Models with a postfix bh0 do not initially possess a central BH,
and are similar to those in \citet{pin95}.
Models with the postfix bh7 or bh8 have a
BH with mass $\MBH(0)=4\times 10^7\Msun$ and $4\times 10^8\Msun$
respectively; they are analogous to models in \citet{mac04b} and
\citet{ann05}.
For most models, we fix the BH mass to its initial value, but we also
consider three additional models (cs05bh0t, cs20bh0t, and
cs20bh7t) in which the BH mass is varied with time according to
$\MBH(t) = \MBH(0) + \int_0^t \Mdot(t^\prime)dt^\prime$,
where $\Mdot$ is the mass inflow rate across the inner boundary (see below).
These time-varying $\MBH$ models allow us to study the effect of BH
growth due to the gas accretion on bar substructures.

\begin{figure}[!t]
\epsscale{1.1} \plotone{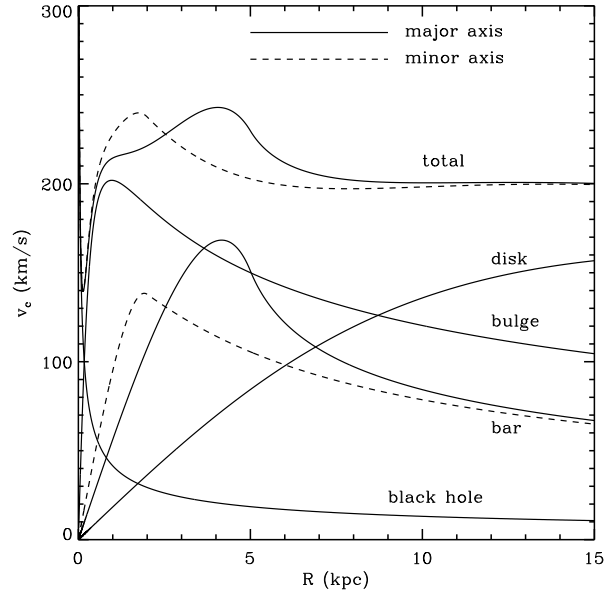}
\caption{Rotational velocity of each component of the model galaxy with a central BH of
$\MBH=4\times10^8\Msun$.
The solid and dashed lines are for along the bar major and minor axes,
respectively.  Note that the effect of the BH is almost
negligible at $R>1\kpc$,
while it dominates the total gravitational potential at $R\simlt 0.1\kpc$.
\label{fig:rotcurve}}
\vspace{0.2cm}
\end{figure}

Figure \ref{fig:rotcurve} plots the net circular rotation curves
together with a contribution from each component when $\MBH=4\times10^8\Msun$.
The solid and dashed lines are along the bar major and minor axes,
respectively.  The circular velocity is almost flat at $\sim 200\kms$
in the outer parts.
Without the BH, the rotation curve $v_c$ would rise linearly
with $R$ close to the center, but the presence of the BH results in
$v_c \propto R^{-1/2}$ for $R\simlt 0.1\kpc$.  This rapid increase
of $v_c$ will render the gaseous orbits in the very central regions highly
resistant to pressure perturbations, resulting in smaller mass inflow
rates than the cases without it, as we show below.

\begin{figure}
\vspace{-0.5cm}
\epsscale{1.2} \plotone{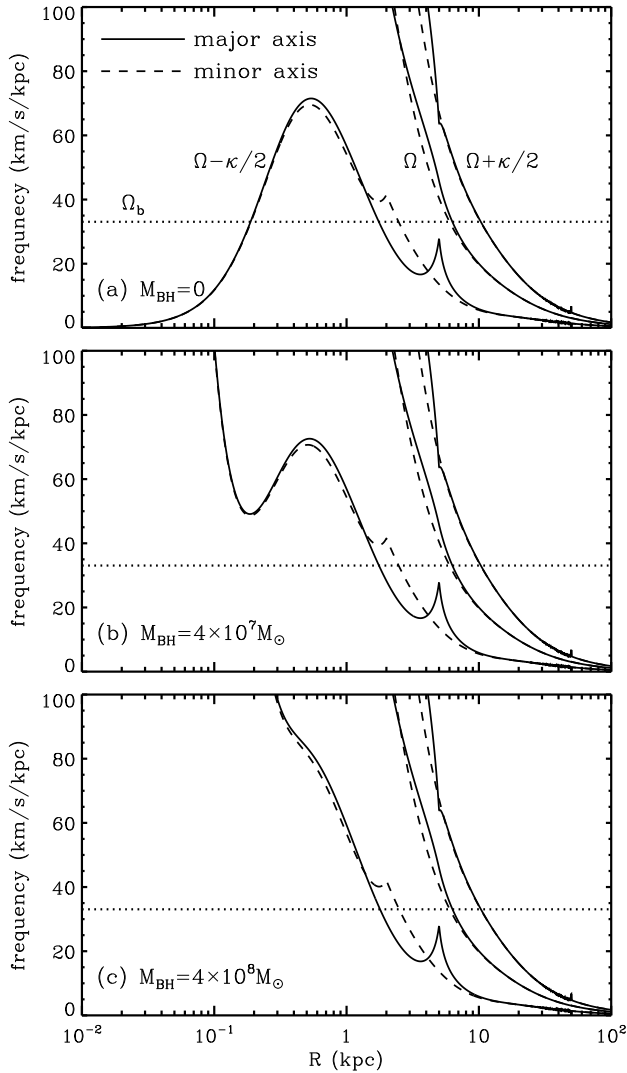}
\caption{Angular frequencies
of galaxy models with different BH masses. The solid and dashed
lines represent $\Omega-\kappa/2$ (rightmost curves), $\Omega$
(middle curves), and  $\Omega+\kappa/2$ (leftmost curves) along the
bar major and minor axes, respectively.  The dotted lines denote the
bar pattern speed $\Omega_b$.  (a) Models without BH have two ILRs
at $\IILR=0.19\kpc$ and $\OILR\approx2\kpc$, with the maximum of the
$\Omega-\kappa/2$ curve occurring at $\Rmax=0.53\kpc$. (b) Models
with $\MBH=4\times10^7\Msun$ have a single ILR at
$\ILR\approx2\kpc$, with the local maximum and minimum of the
$\Omega-\kappa/2$ curve occurring at $\Rmax=0.53\kpc$ and
$\Rmin=0.19\kpc$. (c) Models with $\MBH=4\times10^8\Msun$ have a
single ILR at $\ILR\approx2\kpc$, with $d(\Omega-\kappa/2)/dR<0$ in
the nuclear regions with $R<1\kpc$. \label{fig:angfreq}}
\vspace{-0.3cm}
\end{figure}

\begin{figure*}
\vspace{-0.5cm}
\epsscale{1.0} \plotone{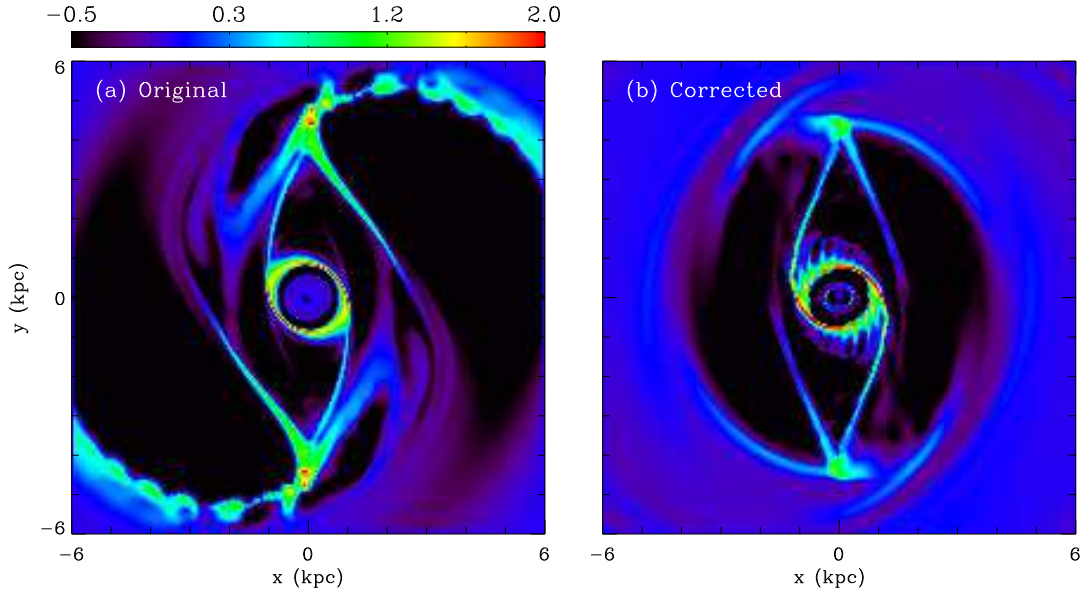} \caption{Logarithm of the
gas surface density at $t = 300$ Myr from a test run using (a) the
original CMHOG code and (b) the corrected version used in this work.
A cylindrical grid with $251\times 154$ zones is used. Compared to
the left panel, gas in the right panel is relatively featureless in
the corotation region at $R = 6\kpc$, has a more clumpy ring, and
harbors nuclear spirals in the central region.
\label{fig:comp_ori_new}}
\end{figure*}

Figure \ref{fig:angfreq} shows the characteristic angular
frequencies, $\Omega-\kappa/2$, $\Omega$, and $\Omega+\kappa/2$
along the bar major and minor axes as solid and dashed lines,
respectively.\footnote{In the presence of a non-axisymmetric bar
potential, the concepts of $\Omega$ as an angular frequency and
$\kappa$ as a radial frequency do not apply strictly since closed
orbits are in general non-circular. In our models, however, the bar
potential is nearly axisymmetric at $R<1\kpc$, so that $\Omega$ and
$\kappa$ measure the actual frequencies reasonably well in the
central parts.} Here, $\Omega^2\equiv R^{-1}d\Phi_{\rm ext}/dR$ and
$\kappa^2\equiv R^{-3}d(R^4\Omega^2)/dR$ denote the angular and
epicyclic frequencies, respectively. The horizontal dotted line in
each panel represents the bar patten speed of $\Omb=33\freq$, with
the corotation resonance (CR) located at $\RCR=6\kpc$ for all
models. For bh0 models with no BH, the $\Omega-\kappa/2$ curve peaks
at $\Rmax=0.53\kpc$ and is equal to $\Omb$ at the two ILR with radii
of $\IILR =0.19\kpc$ and $\OILR\approx 2\kpc$. Because the rotation
curve rises steeply toward the center, bh7 and bh8 models have only
a single ILR at $\ILR\approx2\kpc$. The $\Omega-\kappa/2$ curve in
bh7 models attains its local maximum and minimum at $\Rmax=0.53\kpc$
and $\Rmin=0.19\kpc$, respectively. In bh8 models, on the other
hand, it increases monotonically with decreasing $R$ since the BH
dominates the gravitational potential. We will show in Section
\ref{sec:nsp} that the shape of nuclear spirals depends critically
on the sign of $d(\Omega-\kappa/2)/dR$ (e.g., \citealt{but96}).

\subsection{Numerical Methods}

To solve equations (\ref{eq:con}) and (\ref{eq:mom}), we use the
two-dimensional grid-based code CMHOG in cylindrical geometry
\citep{pin95}. CMHOG implements the piecewise parabolic method in
its Lagrangian remap formulation \citep{col84}, which is third-order
accurate in space and has very little numerical diffusion
(viscosity). All the runs are carried out in a frame corotating with
a bar whose major axis is aligned along the $y$-axis (i.e.,
$\phi=\pm\pi/2$), so that the bar potential remains stationary in
the simulation domain. By assuming a reflection symmetry with
respect to the galaxy center, the simulations were performed on a
half-plane with $-\pi/2\leq\phi\leq\pi/2$ constructed by making a
cut along the bar major axis.

As mentioned in Section \ref{sec:intro}, the original version of CMHOG
used by \citet{pin95} contained a serious bug in the way the gravitational
forces were added to the hydrodynamic equations.
The CMHOG code places a bar potential $\Pbar$ with
the major axis aligned along the $x$-axis, calculate the bar forces
$\fx=-\partial \Pbar/\partial x$ and $\fy=-\partial\Pbar/\partial y$,
and then transforms them into cylindrical coordinates.  In the original version
of the code, the
incorrect transform relations $\fr  =\fx\cos\phi + \fy\sin\phi$ and
$\fphi  = \fx\sin\phi - \fy\cos\phi$ were used.  In fact, the correct
transformation
rule for the azimuthal force should be $\fphi  =-\fx\sin\phi + \fy\cos\phi$.
With the sign of
the azimuthal force reversed (but the radial force correct), the flows in
models computed using the original
CMHOG code behave as if the bar potential were aligned parallel to
the $y$-axis, but with forces quite different from
the intended ones.  Other than these force transformations,
the complex hydrodynamic algorithms in CMHOG are
all implemented correctly, and were well tested in the original paper
by \citet{pin95}.
Therefore, previous numerical studies based on CMHOG should remain valid
as long as they did not adopt the incorrect transformations of the bar forces
inherited from \citet{pin95}.
Unfortunately, the results of \citet{pin95} were compromised by a trivial
sign error in the coordinate transform relations for the bar forces.

\begin{figure*}
\epsscale{1.1} \plotone{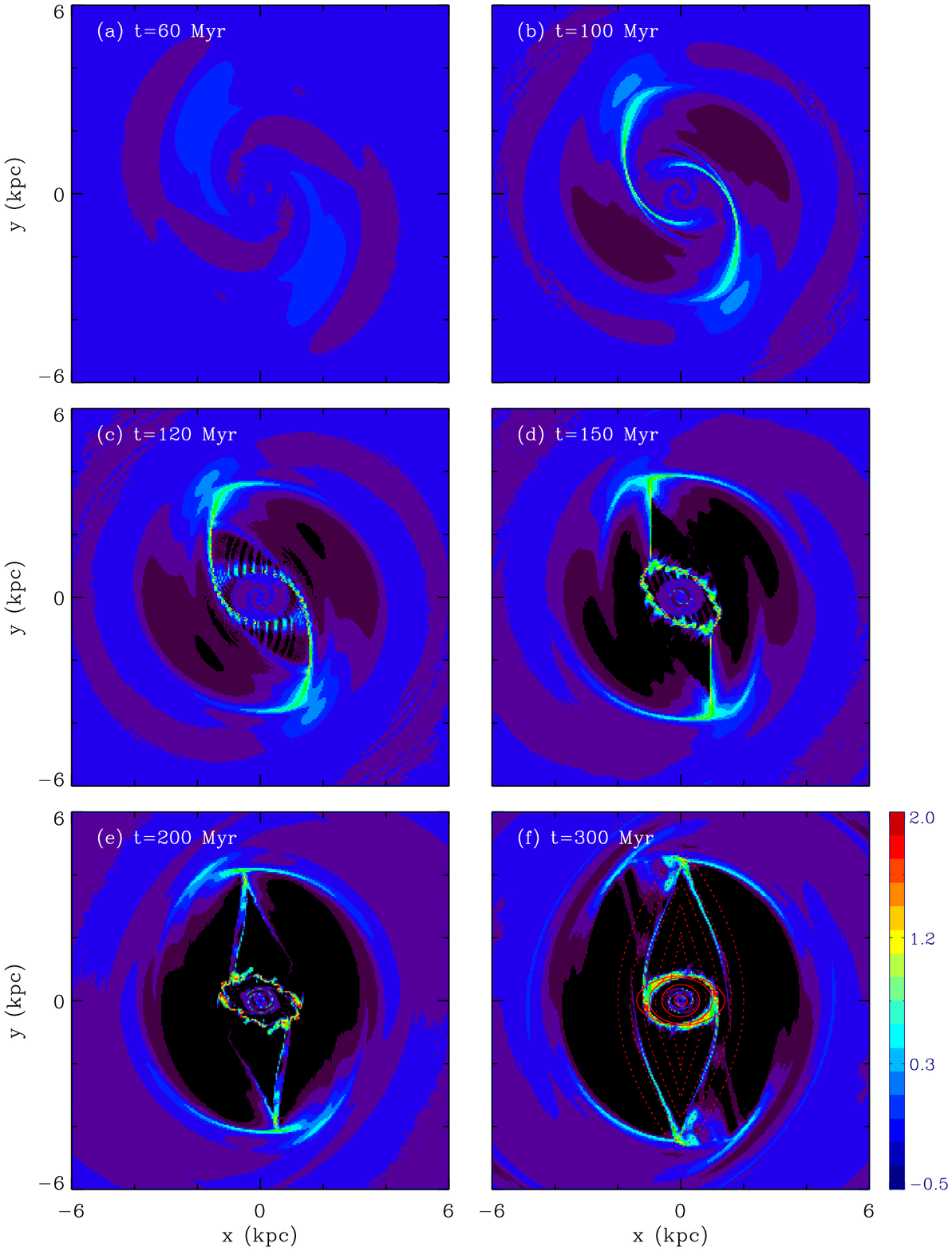} \caption{ Snapshots of the
logarithms of the gas surface density of Model cs05bh0. The bar is
oriented vertically along the $y$-axis and remains stationary. The
gas inside the CR is rotating in the counterclockwise direction.  In
(f), the dotted curves aligned vertically represent the
$\xone$-orbits that cut the $x$-axis at $x_c=0.4$, 0.8, 1.2,
$1.6\kpc$ from inside to outside, while the solid curves aligned
horizontally plot the $\xtwo$-orbits with $x_c=0.2$, 0.6, 1.0,
$1.4\kpc$. Clumpy structures in (c)-(e) are produced by vortex
generation at the curved shocks. See text and Figure \ref{fig:pvort}
for detail. \label{fig:snap}}
\end{figure*}

Figure 3 compares the results for a typical simulation run with the
original and corrected version of CMHOG. For this test, the grid
resolution is taken identical to that in \citet{pin95} with 251 and
154 zones in the radial and azimuthal directions, respectively. The
figure plots the logarithm of the surface density at $t = 300$ Myr.
Several differences are apparent. For example, the gas around the
corotation resonance (CR) at $R = 6\kpc$ in the left panel is
largely evacuated and has corrugated streams linked to the ends of
the bar major axis, whereas the CR region is relatively featureless
in the right panel. The nuclear ring in the left panel is fairly
smooth, while it is quite clumpy in the right panel.  Most
importantly, the very central region inside the ring is almost
unperturbed in the left panel, while the right panel shows spiral
structures in the central region. These differences suggest that the
original CMHOG code is unable to properly model the flow in the
central regions, especially weak nuclear spirals.  We have also run the
same model using other grid-based codes adopting Cartesian coordinates,
such as TVD \citep{kim99} and ANTARES \citep{yua05} as well as the
particle-based GADGET code \citep{spr01}, in all of which the bar
forces are calculated by taking finite differences of $\Pbar$ rather
than using $\fx$ and $\fy$ directly.  The new version of the CMHOG
code used in this work gives results which are much more similar to
the results of these other codes, which gives us further confidence that
the gravitational forces due to the bar are now being treated
correctly.

To resolve the central regions accurately, we set up a non-uniform,
logarithmically-spaced cylindrical grid with 1024 radial zones
extending from $0.02\kpc $ to $16\kpc$ and 480 azimuthal zones
covering the half-plane. This makes the zones approximately
square-shaped throughout the grid (i.e., $\Delta R =R\Delta \phi$).
The resulting grid spacing is $\Delta R=0.13$, $6$, and $100\pc$ at
the inner radial boundary, $R=1\kpc$ where nuclear rings typically
form, and the outer radial boundary, respectively. This increases
the resolution in the inner regions by over an order of magnitude,
in comparison to the models presented in \citet{pin95}. This level
of grid resolution is crucial to resolve nuclear spiral structures
within $R=1\kpc$. We use outflow and continuous boundary conditions
at the inner and outer radial boundaries, respectively, while the
azimuthal boundaries are periodic. The gas crossing the inner
boundary is considered lost from the simulation domain.  We keep
track of the total mass crossing the inner boundary in order to
study the mass inflow rates into the galactic nucleus.

Each model starts from a uniform disk with surface density
$\Sigma_0=10\Surf$ that is rotating in force balance with an
axisymmetric gravitational potential without a bar. In order to
avoid strong transients in the fluid flow caused by a sudden
introduction of the bar, we slowly introduce the bar potential over
one bar revolution time of $2\pi/\Omega_b=186$ Myr. This is
accomplished by increasing the bar central density $\rhobar$
linearly with time and decreasing the bulge central density
$\rhobul$, while keeping the net central density $\rhobar+\rhobul$
fixed.  This ensures that the shape of the total gravitational
potential $\Phi_{\rm ext}$, when averaged along the azimuthal
direction, is unchanged with time. All the models are run until 500
Myr. This corresponds to $1.2\times10^4$ and 10 orbits at the inner
and outer radial boundaries, respectively, for bh8 models with
$\MBH=4\times10^8\Msun$.

\begin{figure}
\epsscale{1.0} \plotone{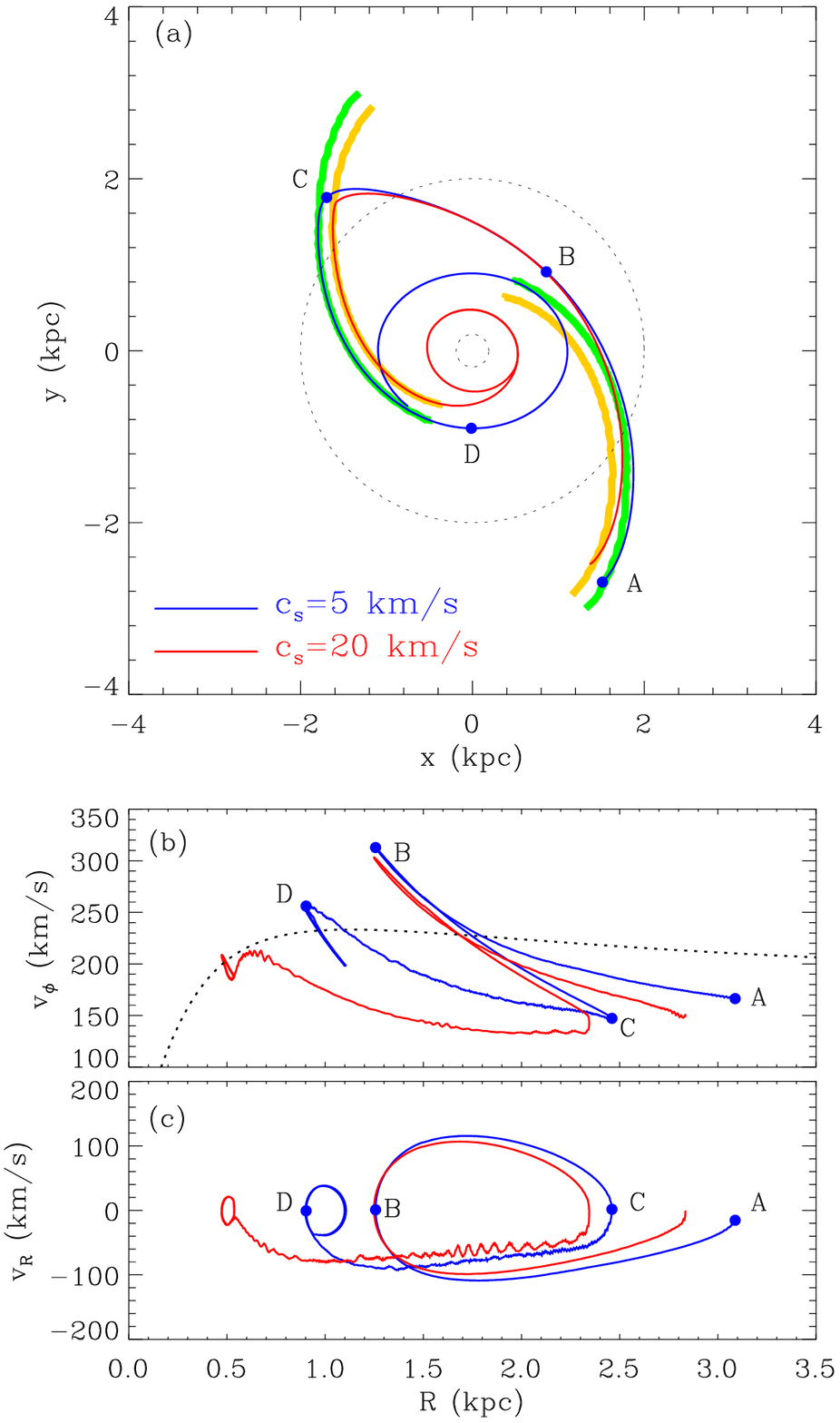} \caption{(a)
Instantaneous streamlines of gas that starts from Point A
$(x,y)=(1.5, -2.7)\kpc$ in Model cs05bh0 and $(1.4, -2.5)\kpc$ in
Model cs20bh0 at $t=100$ Myr. The thick green and orange curves
represent the overdense ridges in Models cs05bh0 and cs20bh0,
respectively.  The two dotted circles indicates $\IILR=0.19\kpc$ and
$\OILR=2\kpc$. In Model cs05bh0, the gas reaches Point B closest to
the galaxy center, is shocked at Point C, and forms a ring at Point
D. (b,c) The variations of the azimuthal and radial velocities of
the gas along the paths shown in (a).  The initial equilibrium
circular velocity is shown as a dotted line in (b).
\label{fig:stream}}
\end{figure}

\section{Results}

We take Model cs05bh0 with $\cs=5\kms$ and no BH as our standard model.
The overall evolution of other models with different $\cs$ and $\MBH$ are
qualitatively similar, although the properties of the nuclear features
that form differ considerably from model to model.
In this section, we first describe the evolution of our standard model,
and then present the differences in the off-axis shocks and
nuclear rings caused by differing $\cs$ and $\MBH$.
The properties of nuclear spirals will be given in Section \ref{sec:nsp}.

\subsection{Overall Evolution}\label{sec:evol}

Figure \ref{fig:snap} plots snapshots of the logarithm of the gas
density at a few selected epochs in the inner regions of Model
cs05bh0. The bar is oriented vertically along the $y$-axis, and the
gas is rotating in the counterclockwise direction relative to the
bar. The images extend to 6 kpc on either side of the center,
corresponding to the CR radius, outside of which the gas remains
almost unperturbed\footnote{The non-axisymmetric bar potential we
adopt is very weak at $R>\RCR$.}. \citet{pin95} found that the CR
regions exhibit time-dependent flow structures, as reproduced in
Figure \ref{fig:comp_ori_new}a. On the other hand, Figure
\ref{fig:snap} shows that the CR regions in our simulations are
quite stable and exhibit only at late time low-amplitude wavelike
features entrained by the dense gas located at the bar ends, similar
to the results of SPH simulations (e.g., \citealt{eng97,pat00}).
This indicates that the complicated structures seen in \citet{pin95}
were likely an artifact of the errors in their force transformation.

\begin{figure*}
\vspace{-0.3cm}
\epsscale{1} \plotone{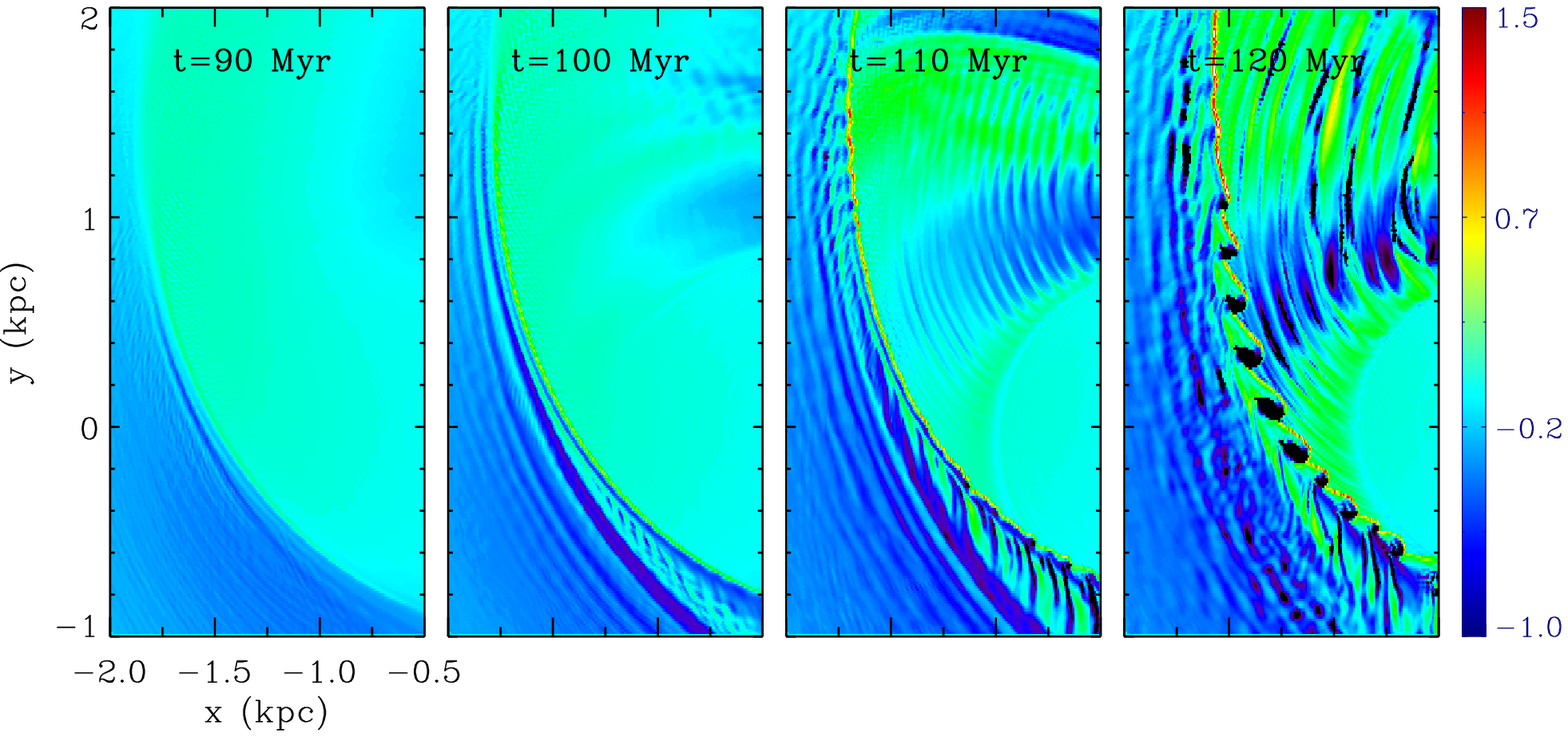} \caption{ Snapshots of the
potential vorticity $\xi\equiv |\nabla\times \mathbf{u} +
2\mathbf{\Omega}|/\Sigma$ normalized by the initial value $\xi_0$ in
Model cs05bh0. Only the regions with $-2\kpc\leq x \leq-0.5\kpc$ and
$-1\kpc\leq y \leq2\kpc$ around the off-axis shocks are shown.
Colorbar shows $\log (\xi/\xi_0)$. This vortex-generating
instability of curved shocks appears to be similar to the wiggle
instability of spiral shocks identified by \citet{wad04}.
\label{fig:pvort}}
\end{figure*}

A striking feature of each of the snapshots shown in Figure
\ref{fig:snap} at times greater than 120 Myr is large amplitude
oscillations in the density in the ring and dust lanes.  These
features are also seen in the simulations of \citet{wad04} for
spiral shocks, and are attributed to the ``wiggle instability". As
we will discuss below, this shock instability appears to be caused
by vorticity generation in curved shocks.

An introduction of the non-axisymmetric bar potential induces perturbations
on the gas orbits, causing them to deviate from circular trajectories.
The gas density increases (decreases) in regions where neighboring orbits
come close together (diverge).  When $t=60$ Myr,
the overdense regions are preferentially located downstream from the bar
major axis (Fig.\ \ref{fig:snap}a).  At this time, the overdensity produced by
orbit crowding is largest at $R\sim3\kpc$.  Since the perturbing force
by the bar is weak inside $R\sim1\kpc$, the overdensity there is
correspondingly small and not readily discernible.
Over time, the overdense regions become narrower and sharper as
the bar amplitude grows and eventually develop into shock fronts at around
$t=100$ Myr. In what follows, we term these narrow shocks off-axis shocks.

A nuclear ring is beginning to shape at this time, as well.
To illustrate the formation of nuclear rings in our models, we plot
as solid lines in Figure \ref{fig:stream}
instantaneous streamlines of the gas that starts from Point A marked
at $(x,y)=(1.5, -2.7)\kpc$ in Model cs05bh0 and
from $(x,y)=(1.4, -2.5)\kpc$ in Model cs20bh0 with $\cs=20\kms$ and no BH
at $t=100$ Myr.
The two dotted circles in Figure \ref{fig:stream}a indicate
the inner and outer ILRs at $\IILR=0.19\kpc$ and $\OILR=2\kpc$.
Note that the thick lines representing the overdense ridges in both models
directly cross the outer ILR.
The changes of the azimuthal and
radial velocities along the streamlines are shown in Figure
\ref{fig:stream}b,c, where  the dotted line indicates the
equilibrium rotation curve of the model galaxy with no BH.
On emerging from the overdense region (Point A), the gas
moves radially inward on its epicycle orbit and increases
(decreases) its azimuthal (radial) velocity due to the Coriolis force.
It reaches Point B closest to the center when it attains $\vR=0$
and largest $\vphi$.  After this point, it moves radially
outward, decreasing $\vphi$ until it hits the off-axis shock at Point C.
The gas loses a significant amount of angular momentum at the shock
and begins to fall in.
In addition, the shocked gas is swept by other shocked gas
flowing from the bar end regions along the shocks.
Note that the shape of the off-axis shocks shown in Figure
\ref{fig:stream}a coincides with the gas streamline from Points C to D,
indicating that all the gas after crossing the shocks moves
radially in along the shock fronts in the developing stage of
the nuclear rings.

As the shocked gas moves along the shock fronts from Point C, it
gradually rotates faster again. When the azimuthal velocity of the
gas is increased to the level comparable to the equilibrium circular
velocity at some radius $R$, it begins to follow a closed orbit
(Point D), forming a nuclear ring at that radius.  In other words,
the centrifugal barrier inhibits the inflowing gas from moving
further in. Regardless of the BH mass, this happens at
$R\sim(0.8-1.2)\kpc$ in models with $\cs=5\kms$ and
$R\sim0.4-0.6\kpc$ in models with $\cs=20\kms$
(Fig.\ \ref{fig:snap}b,c). The facts that the off-axis shocks penetrate the
ILR in bh7/bh8 models, the outer ILR in bh0 models, and that the
ring positions are almost independent of $\MBH$
when the rings are beginning to form suggest that the ring
formation is unlikely to be governed by resonances. For models with
low sound speed, the shape of nuclear rings is similar to an
$\xtwo$-orbit. Clearly, the presence of the nuclear ring prevents
the shocked gas from infalling directly to the nucleus.

Since the off-axis shocks are curved, Crocco's theorem ensures that
vorticity can be generated at the shock fronts. Figure \ref{fig:pvort}
plots snapshots of the potential vorticity
$\xi\equiv |\nabla\times \mathbf{u} + 2\mathbf{\Omega}|/\Sigma$
relative to the initial value $\xi_0$
near the shocks at
$t=90$, 100, 110, and 120 Myr of the standard model.
At $t=90$ Myr, $\xi/\xi_0$ is largest along the shocks.
Vorticity produced at the shocks is advected with the background
flows and enters the shock fronts at the opposite side after a half
revolution.  Vorticity grows secularly with time by
successive passages across the shocks.
When vorticity achieves substantial
amplitudes, it causes the shock fronts to wiggle and fragment
into small clumps with high vorticity (see Fig.\ \ref{fig:snap}c).
The process of clump formation along the shocks in our models bears
remarkable resemblance to the wiggle instability of spiral shocks found
by \citet{wad04} (see also \citealt{kim06}).  These clumps are carried
radially inward and add to the nuclear ring, making the latter
fairly inhomogeneous (Fig.\ \ref{fig:snap}d,e).

\begin{figure*}
\vspace{-1.0cm}
\epsscale{1.1} \plotone{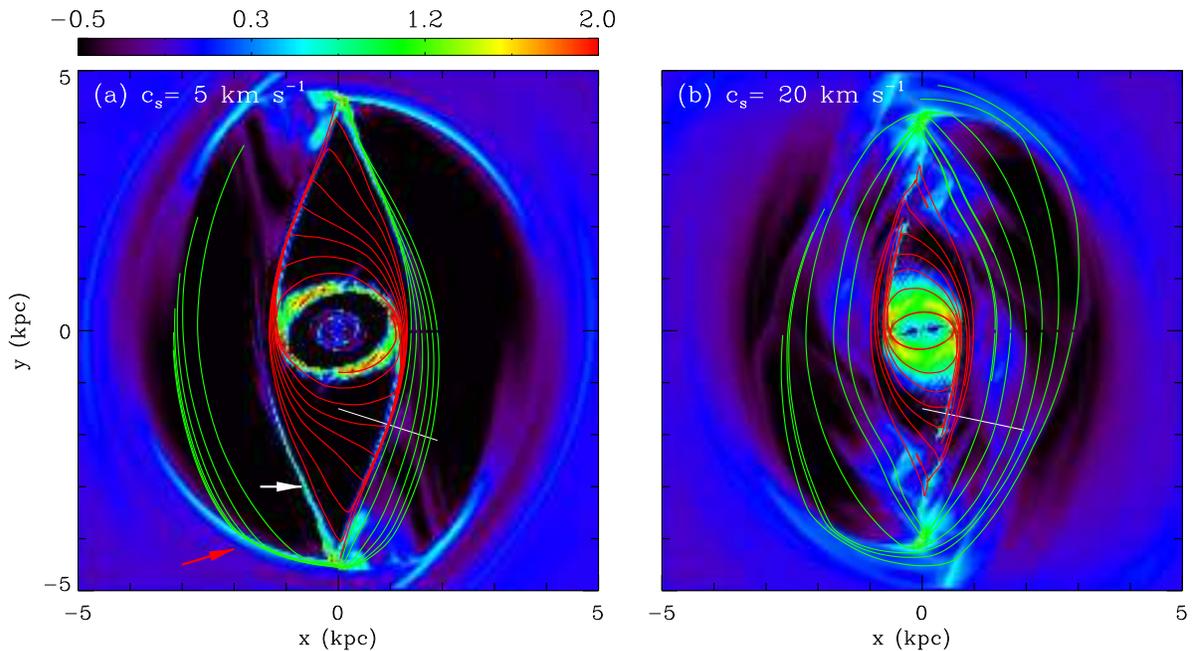} \caption{Logarithm of the density
distribution overlaid with instantaneous
streamlines in Models cs05bh0 and cs20bh0 at $t=300$ Myr. The red
lines represent streamlines that meet the off-axis shocks, while the
green lines are for those that go around the shocks.  The red and
white arrows in (a) mark the 4/1-spiral shocks and ``smudge",
respectively.  The short white line segment in each panel indicates
a slit along which density and velocity are measured in Figure
\ref{fig:slit}. \label{fig:csVar}}
\end{figure*}

The off-axis shocks shown in Figure \ref{fig:stream} are not stationary
largely because the bar potential is not fully turned on yet.
As the strength of the bar potential keeps increasing, they become stronger
and move slowly toward the bar major axis.
Gas that is added to the ring from the off-axis shocks has
increasingly lower angular momentum with time, causing
the ring to shrink in radius with time.
After the bar potential is fully turned on, the off-axis shocks become
gradually weaker as the amount of gas in the mid-bar regions lost to
the ring increases with time.
At the same time, orbital phase mixing and frequent clump collisions
in the ring make the latter rounder and align its semimajor axis
in the direction perpendicular to the bar major axis.
Note that the rings are always attached to the inner end of the off-axis
shocks.

At $t=300$ Myr, Model cs05bh0 reaches a quasi-steady state in the
sense that temporal changes in the overall flow pattern are very
slow, and the locations of the shocks and rings do not vary much with time.
Figure \ref{fig:snap}f overplots some of the $\xone$-orbits (dotted
curves aligned vertically) and $\xtwo$-orbits (solid curves aligned
horizontally), showing that the shape of the nuclear ring matches
well with an $\xtwo$-orbit, while the off-axis shocks closely follow
one of the $\xone$-orbits over the whole length of the shocks. This
is because when $\cs=5\kms$ the impact of thermal pressure on the
gas orbits is much smaller than that of the gravitational and
centrifugal forces, so that pure orbit theory (neglecting pressure
forces) is a good description. When $\cs\simgt 15\kms$, however,
thermal pressure gradients strongly affect gas orbits, and thus the
morphology of substructures in the central regions are modified, as
we will discuss below.

\subsection{Off-axis Shocks}

\begin{deluxetable*}{lcccrrcc}
\tabletypesize{\footnotesize} \tablewidth{0pt}
\tablecaption{Properties of Off-axis Shocks\label{tbl:shock}}
\tablehead{
\colhead{Model}                       & \colhead{$s_{\rm sh}$}  &
\colhead{$\Sigma_{\rm max}/\Sigma_0$} & \colhead{$\Mperp$} &
\colhead{$\Mpara$}                    & \colhead{$\ish$} &
\colhead{$d\vpara/ds$}                & \colhead{$\alpha_{\rm max}$} \\
\colhead{ } & \colhead{(kpc)} & & & & \colhead{(deg)} &
\colhead{$(10^3\kms\kpc^{-1})$} & \colhead{ }
} \startdata
cs05bh0  & 0.98  & 5.7 & 12.1 & 10.9 & 47.7 & 2.7 & 7.2\\
cs10bh0  & 0.72  & 3.6 &  6.0 &  4.9 & 50.8 & 1.8 & 3.5\\
cs15bh0  & 0.55  & 5.7 &  5.7 &  1.0 & 80.4 & 2.3 & 2.6 \\
cs20bh0  & 0.45  & 5.0 &  3.5 &  1.2 & 71.3 & 1.8 & 1.6 \\
\hline
cs05bh7  & 0.93 & 5.8 & 15.4 & 6.3    & 67.9    & 1.8 & 7.4\\
cs10bh7  & 0.63 & 9.8 & 8.1 & $-1.3$ & $-81.4$ & 1.6  & 4.7 \\
cs15bh7  & 0.47 & 9.0 & 6.6 & $-1.2$ & $-79.8$ & 1.7  & 3.0 \\
cs20bh7  & 0.31 & 5.7 & 2.3 & 1.3    & 59.4    & 1.0  & 1.9 \\
\hline
cs05bh8  & 0.96  & 6.1  & 16.7  & 7.2    & 66.7    & 2.0 & 7.5 \\
cs10bh8  & 0.74  & 7.4  & 9.6  & $-0.1$  & $-89.5$ & 2.3 & 6.3\\
cs15bh8  & 0.43  & 6.8  & 7.2  & $-0.8$ & $-83.3$ &1.4  &  3.0\\
cs20bh8  & 0.32  & 21.6 & 5.9 & $-2.6$ & $-66.6$ & 1.4 &   2.9
\tablecomments{ $s_{\rm sh}$ is the shock position along the slit;
$\Sigma_{\rm max}$ is the maximum density attained immediately after
$s_{\rm sh}$; $\Mperp$ and $\Mpara$ are the Mach numbers of the
incident flow perpendicular and parallel to the shock, respectively;
$\ish$ is the inclination angle of the incident flow with relative
to the shock; $d\vpara/ds$ is the velocity shear in the postshock
region;
$\alpha_{\rm max}$ is the maximum value (occurring at the shock
front) of the compression factor $\alpha\equiv-(\nabla\cdot\mathbf{v})
\Delta R/\cs$.}
\end{deluxetable*}

Even if the gravitational potential is the same, the flow morphology
and velocity fields differ considerably depending on the sound speed.
Figure \ref{fig:csVar} shows instantaneous streamlines
plotted over the logarithm of the density distribution in Models
cs05bh0 and cs20bh0 at $t=300$ Myr. Red curves denote the streamlines
that go through the off-axis shocks, while those enveloping the
off-axis shocks are represented by green curves.
In all models, the off-axis shocks are almost parallel to $\xone$-orbits.
They start from the bar major axis at the
outer ends, offset toward downstream in the mid-bar regions, and connect
to the nuclear rings at the inner ends.  The mean offset
of off-axis shocks from the bar major axis is larger for models with
smaller $\cs$.

The outer end regions of the off-axis shocks have complicated
density structures including the ``4/1-spiral shocks'' marked with a
red arrow in Figure \ref{fig:csVar}a \citep{eng97} and the enhanced
density ridges (a white arrow) termed ``smudges'' by \citet{pat00}.
As discussed by \citet{eng97}, the 4/1-spiral shocks are produced by
collisions of gas moving on $\xone$-orbits with that on the
4/1-resonant family (e.g., \citealt{con89}).  The gas loses angular
momentum at the shocks and subsequently switches to lower orbits. As
the streamlines in green display, in models with $\cs=5\kms$,  the
4/1-spiral shocks are quite strong and spatially extended, so that
the orbits after the shocks become relatively radial and converge at
the opposite side of the bar, building a smudge after about a half
revolution. Collisions of streams off the 4/1-spiral shock and the
smudge on the same side of the bar funnel the gas at the
intersections to an $\xone$-orbit, which are the starting points of
the off-axis shocks. When $\cs=20\kms$, on the other hand, the
4/1-spiral shocks are short and weak, and the streamlines off the
shocks diverge, so that
a dense ridge does not form. Since the
gas becomes less compressible with increasing $\cs$, steady off-axis
shocks in models with large $\cs$ can be supported only in inner
regions where the bar perturbations are sufficiently strong.  This
explains why the mean offset of the off-axis shocks from the bar
major axis becomes smaller as $\cs$ increases (e.g.,
\citealt{eng97}). With weak 4/1-spiral shocks and no smudge, the gas
in the bar-end regions in model with $\cs=20\kms$ is comparatively
unsteady, sometimes generating small dense blobs that move inward
along the off-axis shocks.

\begin{figure}
\epsscale{1.2} \plotone{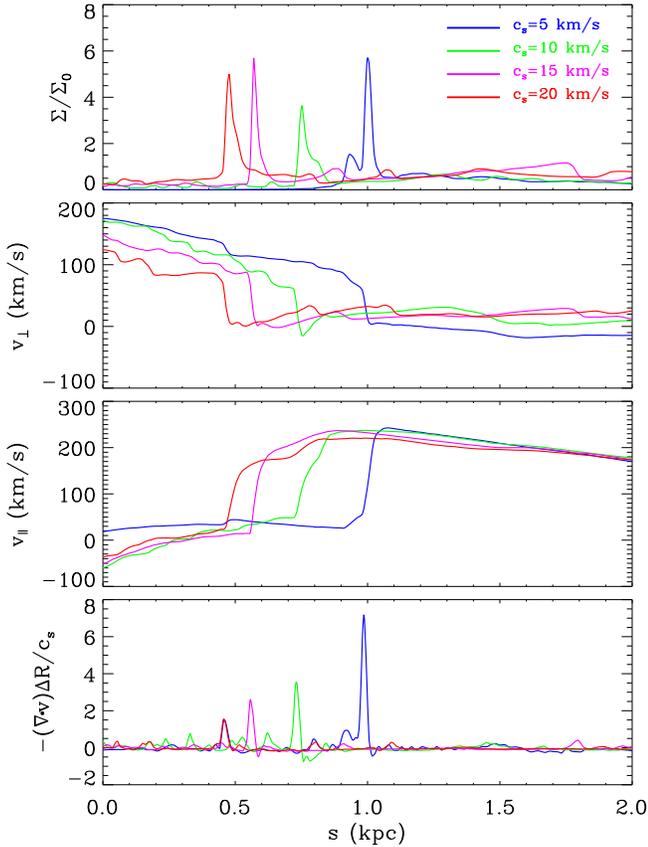} \caption{Profiles of surface
density $\Sigma$, velocity $\vperp$ perpendicular and $\vpara$ parallel to
the off-axis shock, and
compression factor $\alpha=-(\nabla\cdot\mathbf{v})\Delta
R/\cs$
along the slit in bh0 models with differing $\cs$ at $t=300$ Myr.
The position of the slit is shown in Figure \ref{fig:csVar}.
\label{fig:slit}}
\end{figure}

To quantify the shock properties, we place a slit in each model,
indicated by the short white lines in Figure \ref{fig:csVar}.  The slit
starts from $(x,y)=(0, -1.5\kpc)$ and runs perpendicular to the
local segment of the off-axis shocks, roughly at $R\sim(1.5-1.8)\kpc$.
Figure \ref{fig:slit} plots
the profiles along the slit of surface density, velocities,
and the compression factor
\begin{equation}\label{eq:cfac}
\alpha\equiv -(\nabla\cdot \mathbf{v})\Delta R/\cs,
\end{equation}
the last of which can be used as an effective measure
of the shock strength (e.g., \citealt{mac04b, tha09})\footnote{For planar
isothermal shocks in steady state, $\alpha=\Mperp - \Mperp^{-1}$ at the
shock discontinuities.}.
The gas is flowing from left to right in the increasing
direction of $s$,
where $s$ measures the distance along the slit from the starting point.
Table \ref{tbl:shock} gives the shock properties for all
models at $t=300$ Myr: $s_{\rm sh}$ is the
position of the off-axis shocks along the slit,
$\Sigma_{\rm max}$ is the peak density after $s_{\rm sh}$,
$\Mperp\equiv\vperp/\cs$ and $\Mpara\equiv\vpara/\cs$ are the Mach numbers of
the incident flows in the directions perpendicular and parallel
to the shocks, respectively,
$\ish\equiv\tan^{-1}(\Mperp/\Mpara)$ is the inclination angle of the
preshock velocity relative to the shock front,
$d\vpara/ds$ quantifies the velocity shear in the postshock region,
and $\alpha_{\rm max}$ is the maximum value of the compression
factor occurring at the shock front.
It is apparent that the off-axis shocks tend to move toward the bar major
axis with increasing $\cs$, while there is no clear dependence of
$s_{\rm sh}$ on the BH mass.
The compression factor at the shock is insensitive to $\MBH$ and scales
roughly with $\cs$ as $\alpha_{\rm max}\sim 7.7 (\cs/5\kms)^{0.92}$.

\begin{figure*}
\vspace{-1.5cm}
\epsscale{1.0} \plotone{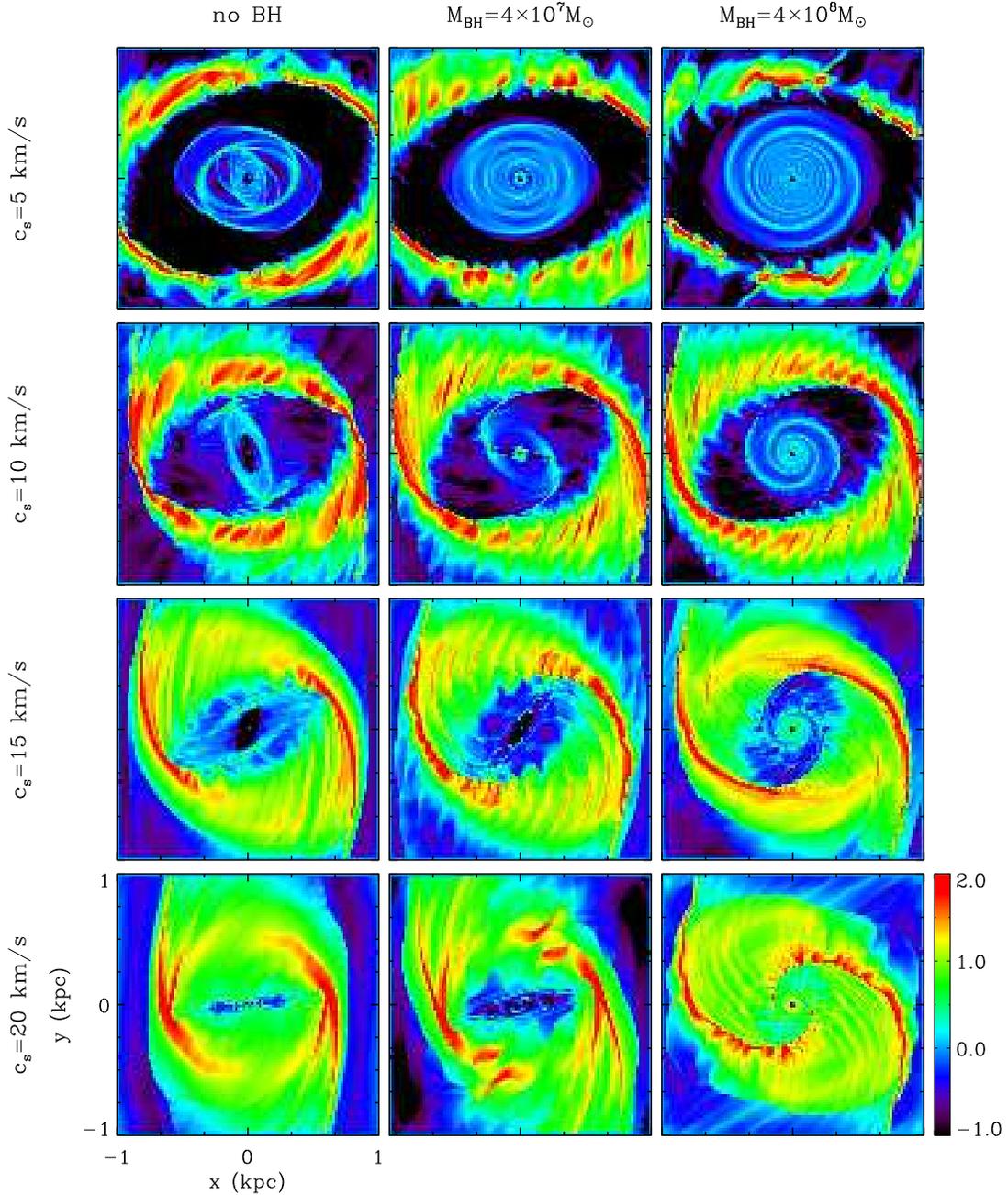}
\caption{Effects of sound speed and BH mass on the
distribution of gas surface density, shown in logarithmic scale,
in the central regions of all models at $t=300$ Myr.
The nuclear rings are narrow when $\cs=5\kms$, while they spread out
as $\cs$ increases.
\label{fig:ringall}}
\end{figure*}

Naively, one would expect that the off-axis shocks become weaker as
the sound speed increases, since the density jump in planer
isothermal shocks is proportional to $\Mperp^2$.  However, this is
not the case, as Figure \ref{fig:slit} and Table \ref{tbl:shock}
demonstrate. There are several reasons for this.  Firstly, it is the
velocity component normal to the shock front $\vperp$ that
determines the shock jump conditions, and because the inclination
angle of the streamlines which enter the shock varies with location
and with $\cs$, $\vperp$ varies in a complicated fashion. For
example, Figure \ref{fig:slit} shows that for off-axis shocks formed
at $R\sim (1.5-1.8)\kpc$, Model cs15bh0 with $\cs=15\kms$ has the
largest peak density as well as the largest $\vperp=85\kms$ and
$\ish=80^\circ$. On the other hand, Model cs10bh0 has the smallest
$\vperp=60\kms$ (with $\ish=51^\circ$) and thus the lowest density
enhancement.  Since the sound speed is lower, Model cs05bh0 with
$\vperp=60\kms$ produces $\Sigma_{\rm max}$ comparable to that in
Model cs15bh0. Secondly, we note that the Rankin-Hugoniot jump
conditions for stationary planar shocks are not applicable to the
curved and two-dimensional off-axis shocks formed in our
simulations.  As Figure \ref{fig:csVar} displays, the flows are
fully two-dimensional in the sense that streamlines diverge before
the shocks, and converge after the shock with the radial inflow
coming from the regions near the end of the bar.  The fact that
the compression factor $\alpha$ measured at the shock front is smaller than
$\Mperp-\Mperp^{-1}$ expected from planer isothermal shocks also indicates
that the shocks are two dimensional.
Finally, the
density and velocity fluctuations generated by the vortex-generating
instability are important around the off-axis shocks especially for
models with small $\cs$, so that the flows are not strictly
stationary\footnote{Negative values of $\vperp$ right after the
shocks in model cs10bh0 shown in Fig.\ \ref{fig:slit} are due to
vortices produced by the instability.}. Note that the shocked gas
has strong velocity shear, amounting to $d\vpara/ds\sim (1-3)\times
10^3\freq$, which is about 10 times larger than the velocity shear
arising from galaxy rotation in the solar neighborhood.  Such strong
shear can stabilize the high-density, off-axis shocks against
self-gravity.

\begin{figure*}
\vspace{-0.5cm}
\epsscale{0.9} \plotone{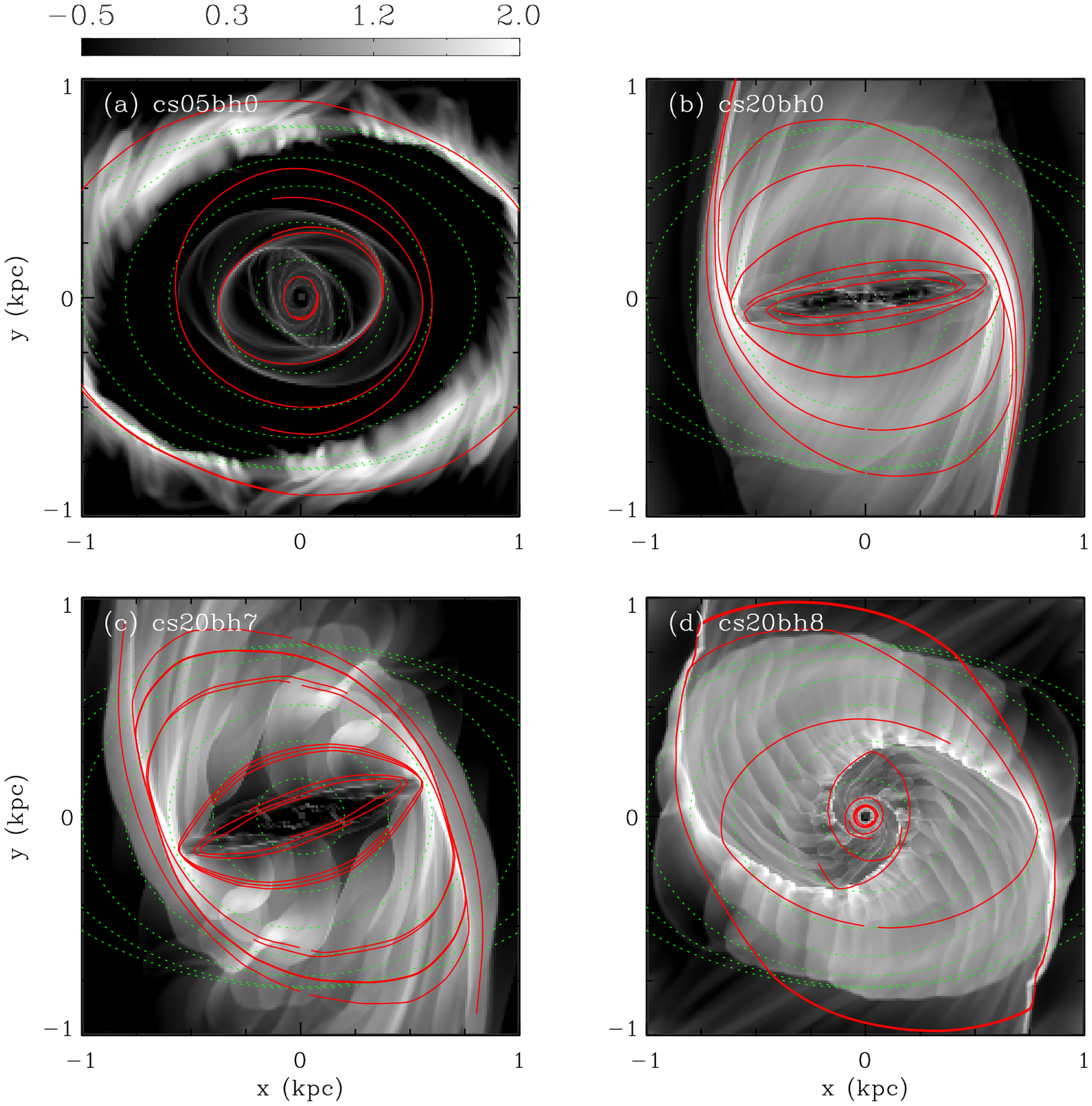}
\caption{Instantaneous streamlines (red solid lines)
overlaid on the logarithm of the density distribution
for models with (a) $\cs=5\kms$ and no BH,
(b) $\cs=20\kms$ and no BH, (c) $\cs=20\kms$ and $\MBH=4\times10^7\Msun$,
and (d) $\cs=20\kms$ and $\MBH=4\times10^8\Msun$, at $t=300$ Myr.
The dotted curves in all panels represent $\xtwo$-orbits.
\label{fig:st_ring}}
\end{figure*}

\begin{figure}
\epsscale{1.1} \plotone{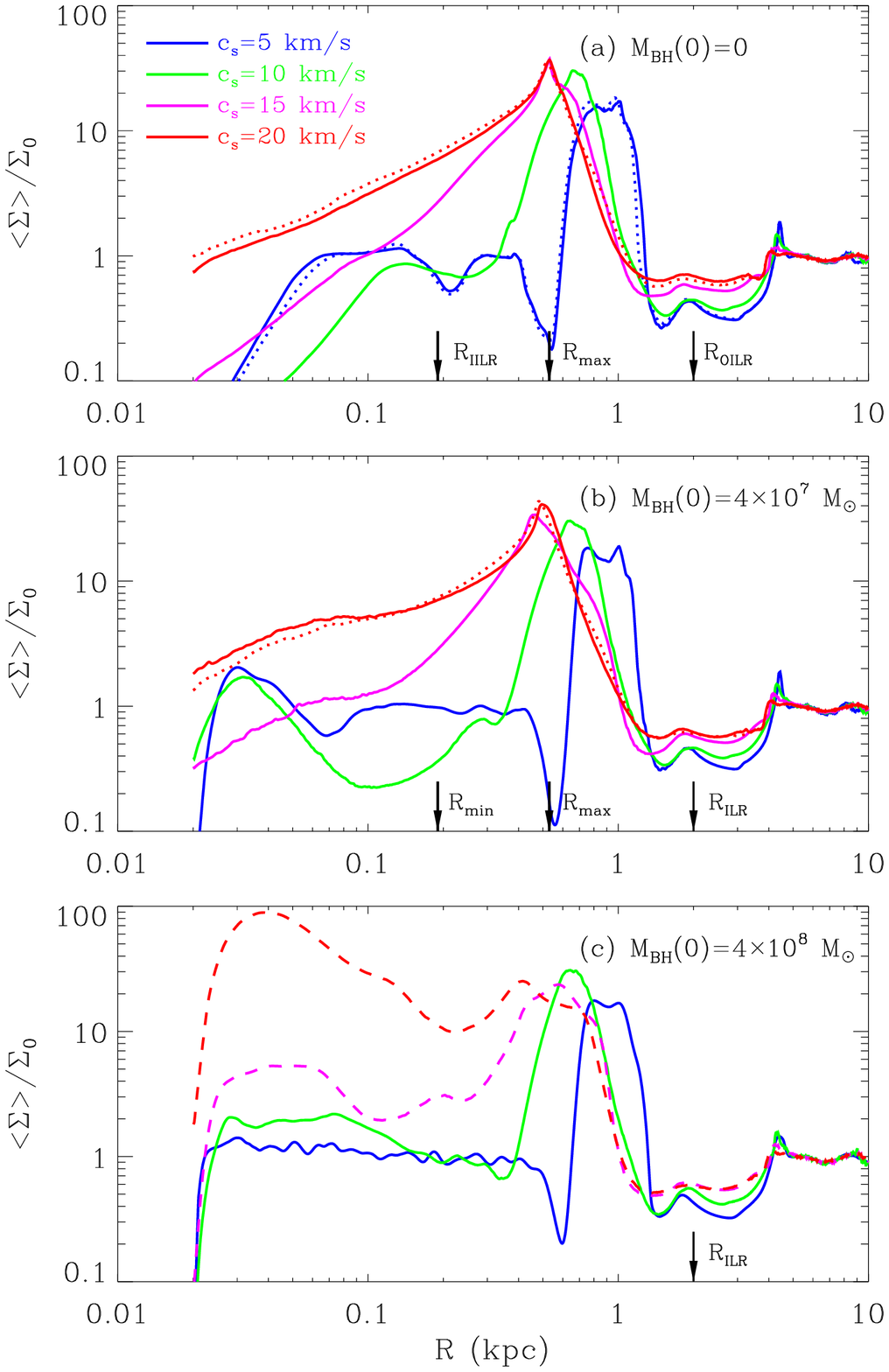}
\caption{
Radial distribution of the gas surface density $\aSig$ averaged both
azimuthally and temporally over $t=300-500$ Myr for models with
(a) $\MBH(0)=0$, (b) $\MBH(0)=4\times10^7\Msun$, and
(c) $\MBH(0)=8\times10^7\Msun$.
The dotted lines are the results of
the models in which $\MBH$ varies with time.
The locations of $\Rmax$ and $\Rmin$ where the $\Omega-\kappa/2$ curve
attains local maximum and minimum and the relevant ILRs
are indicated as arrows along the abscissa.
In (c), the dashed lines correspond to the cases with $\cs=15$ or $20\kms$
for which the density at $R<0.1\kpc$ is dominated by nuclear spirals
rather than rings.
\label{fig:ring_aden}}
\end{figure}

\begin{deluxetable}{lccccc}
\tabletypesize{\footnotesize}
\tablewidth{0pt}
\tablecaption{Properties of Nuclear Rings\label{tbl:ring}}
\tablehead{
\colhead{Model} &
\colhead{$R_{\rm in}$ (kpc)}  &
\colhead{$R_{\rm out}$ (kpc)}  &
\colhead{$R_{\rm ring}$ (kpc)}  &
\colhead{$\aSig_{\rm max}/\Sigma_0$} &
\colhead{$\Sigma_{\rm ring}/\Sigma_0$}
}
\startdata
cs05bh0 &   0.63 &  1.22 &  0.92 &   17.2 &  11.9  \\
cs05bh0t &  0.62 &  1.17 &  0.89 &   18.4 &  13.1  \\
cs10bh0 &   0.45 &  0.90 &  0.68 &   30.2 &  17.9  \\
cs15bh0 &   0.30 &  0.77 &  0.54 &   37.5 &  17.6  \\
cs20bh0 &   0.23 &  0.71 &  0.49 &   36.8 &  16.9  \\
cs20bh0t &  0.20 &  0.73 &  0.48 &   34.7 &  16.1  \\
\hline
cs05bh7 &   0.66 &  1.17 &  0.90 &   18.9 &  13.9  \\
cs10bh7 &   0.46 &  0.88 &  0.67 &   30.3 &  18.9  \\
cs15bh7 &   0.28 &  0.80 &  0.52 &   33.9 &  16.9   \\
cs20bh7 &   0.23 &  0.67 &  0.47 &   41.0 &  19.5  \\
cs20bh7t &  0.22 &  0.66 &  0.46 &   43.4 &  19.9  \\
\hline
cs05bh8 &   0.69 &  1.25 &  0.94 &   18.1 &  12.7  \\
cs10bh8 &   0.47 & 0.89  & 0.67  &   30.9 & 19.4   \\
cs15bh8 & $\cdots$ &  $\cdots$ & $\cdots$  &$\cdots$   & $\cdots$  \\
cs20bh8 & $\cdots$ &  $\cdots$ & $\cdots$  &$\cdots$   & $\cdots$  
\tablecomments{ $R_{\rm in}$ and $R_{\rm out}$ are the inner and
outer radii of the ring defined by the positions where
$\aSig=\aSig_{\rm max}/5$, with $\aSig_{\rm max}$ being the maximum
density; $R_{\rm ring}$ is the mass-weighted ring radius;
$\Sigma_{\rm ring}$ is the mean density of the ring. }
\end{deluxetable}

\subsection{Nuclear Rings}

Gas that loses angular momentum at the off-axis shocks
flows radially inwards and forms a nuclear ring in the central regions.
Figure \ref{fig:ringall} shows diverse morphological features
produced in the regions with  $|x|,|y|\leq 1\kpc$ for all models
at $t=300$ Myr.
Figure \ref{fig:st_ring} overplots instantaneous streamlines
for a few selected models.
Some models (with low $\cs$) have a nuclear ring together with inner
spiral structures,  some models (with high $\cs$ and small $\MBH$)
have a ring with no spirals,
while others (with high $\cs$ and large $\MBH$) possess only
nuclear spirals without an appreciable ring.

When $\cs=5\kms$, the nuclear rings are quite narrow and clearly
decoupled from the nuclear spirals.
Even though the ring has a large density, the low sound speed makes
the effect of thermal pressure on the gas orbits insignificant.
The gas around the ring in Model cs05bh0 thus follows $\xtwo$-orbits
fairly well and the shape of the ring does not deviate considerably
from $\xtwo$-orbits (Fig.\ \ref{fig:st_ring}a).
When $\cs=20\kms$, on the other hand, the pressure force in the central
regions becomes important and affects the shape of the gas streamlines.
Even the inflowing gas that arrives at the contact points
between the off-axis shocks and the nuclear ring takes very different
orbits depending on its location.
Some gas at the outer parts of the contact points is pushed out
by the pressure gradient and follows trajectories that are much rounder
than $\xtwo$-orbits, while the gas in the inner parts is forced
to take inner highly eccentric orbits (Fig.\ \ref{fig:st_ring}b).
Consequently, the gas in the central regions in Models cs20bh0
spreads out spatially and forms a ring that is more circular and
broader than in Model cs05bh0.  Since the presence of a central BH
increases the initial angular momentum of the gas in the central regions,
the pressure effect becomes less important as $\MBH$ increases.
Figure \ref{fig:st_ring} shows that the pressure distortion of
$\xtwo$-orbits is still significant for $\MBH=4\times10^7\Msun$,
while the gas orbits in the very central parts at $R\simlt
0.2\kpc$ remain almost intact when $\MBH=4\times10^8\Msun$.
In Model cs20bh8, some gas at $R\sim 0.5-0.8\kpc$ temporarily moves
radially outward due to the radial pressure gradient built up
by the background gas and is subsequently
swept inward by other gas flowing in along the off-axis shocks.

Figure \ref{fig:ring_aden} plots the radial distribution of the averaged
gas surface density $\aSig$, averaged both azimuthally and temporally
over $t=300-500$ Myr for all models.
The locations of the ILRs as well as $\Rmax$
and $\Rmin$ corresponding to the local maximum and minimum of the
$\Omega-\kappa/2$ curve are indicated by arrows on the abscissa.
Table \ref{tbl:ring} gives the inner and outer radii, $R_{\rm in}$
and $R_{\rm out}$, of the ring, the mass-weighted ring radius
$\Rring=\int_{R_{\rm in}}^{R_{\rm out}} \aSig RdR /
\int_{R_{\rm in}}^{R_{\rm out}} \aSig dR$,
the peak density $\aSig_{\rm max}$,
and the mean density $\Sigma_{\rm ring}=\int_{R_{\rm in}}^{R_{\rm out}}
\aSig dR/(R_{\rm out}-R_{\rm in})$ of the ring in each model. Here,
$R_{\rm in}$ and $R_{\rm out}$ are defined as the radii where
$\aSig=\aSig_{\rm max}/5$.
Note that Models cs15bh8 and cs20bh8 do not harbor a
well-defined nuclear ring,
as will be discussed in the next section.

All rings that form are located within $\OILR$ if there are two ILRs or $\ILR$
if there is a single ILR, indicating that the formation of a nuclear ring
does not require the presence of two ILRs.  However, there is in general no
direct connection between the ring positions and $\OILR$ or $\ILR$.
When $\cs=5\kms$, the rings are all located at $R\sim0.6-1.2\kpc$,
independent of $\MBH$.
The mass-weighted radius is $\Rring\sim(0.90-0.92)\kpc$,
indicating that the ring position is not governed by
the shape of the $\Omega-\kappa/2$ curve.
When $\cs=10\kms$, the ring radius decreases to $\Rring\sim (0.67-0.68)\kpc$,
again insensitive to the BH mass,
consistent with the tendency for the off-axis shocks to move
closer to the bar major axis as $\cs$ increases.
Rings with $\cs=10\kms$ have a larger surface density than those with
$\cs=5\kms$, corresponding approximately to a constant ring mass
(i.e., $\Sigma_{\rm ring}\propto \Rring^{-1}$).
As $\cs$ increases further, high thermal pressure provides
strong perturbations for $\xtwo$-orbits and tend
to spread out the gas in the central parts, resulting in a broad
distribution of $\aSig$ at $R\simlt0.5\kpc$.
When the BH is not massive enough, these perturbations wipe out
coherent, weak spiral structures that formed earlier in the nuclear regions.

Because the presence of a central BH dominates the potential only in the
central region,  the allowance for the growth of the BH due to
mass accretion does not make significant difference in the regions outside
the ring.  Even inside the ring, Figure \ref{fig:ring_aden} and
Table \ref{tbl:ring} show that the changes in the ring size $\Rring$ and
the ring density $\Sring$ caused by the temporal change of $\MBH$ are
less than 4\% and 10\%, respectively.
We will show below that BH growth in our simulations does not significantly
affect the properties of nuclear spirals and mass inflow rates, as well.

\begin{figure*}
\epsscale{0.8} \plotone{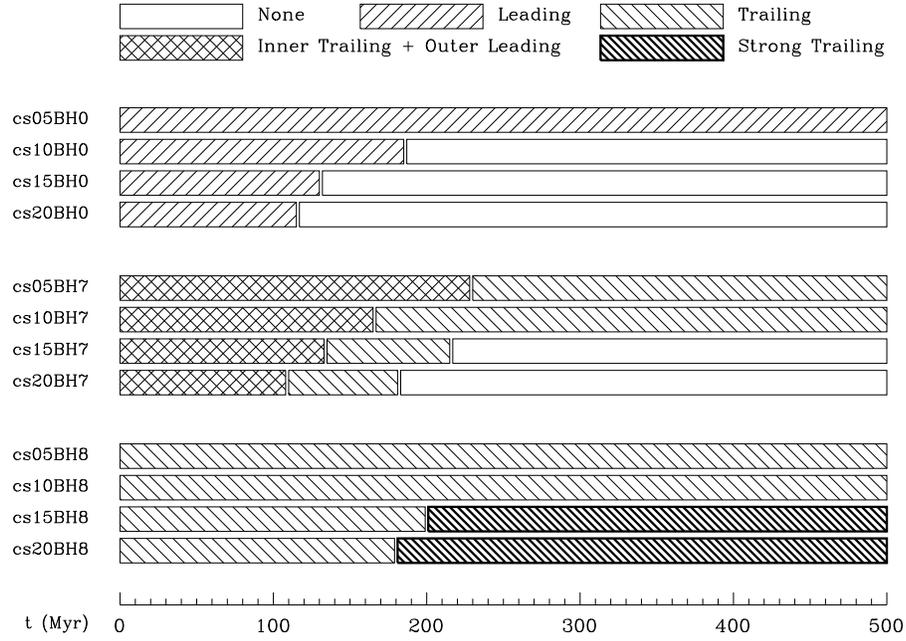}
\caption{
Schematic illustration of the types of nuclear spirals
found in our simulations, and their duration.
\label{fig:nsp_class}}
\end{figure*}

\section{Nuclear Spirals}\label{sec:nsp}

High-resolution observations of barred galaxies reveal that
some contain nuclear spirals inside a nuclear ring
(e.g., \citealt{mar03a,mar03b,pri05,van10}).
As Figures \ref{fig:ringall} displays,
some of our models also have spiral structures in their nuclear regions
that persist for long periods of time.  Other models also have nuclear
spirals at early time but they are destroyed as a result of interactions
with nuclear rings.  In this section, we describe the formation and
shape of nuclear spirals in detail.  Figure \ref{fig:nsp_class}
schematically summarizes how the type of nuclear spirals changes with time
and how long they survive in each model.
Table \ref{tbl:spiral} lists the properties of nuclear
spirals measured at $t=500$ Myr for the models that possess
long-lasting spirals: $R_{\rm in}$ and $R_{\rm out}$ denote the radii
of the inner and outer ends of the nuclear spirals, respectively;
$\bSig$ and $\pSig$ are the azimuthally-averaged and peak surface densities,
respectively, at $R=R_a=0.05\kpc$ for Models cs15bh8 and cs20bh8
and $R_a=\sqrt{R_{\rm in}R_{\rm out}}$ for the other models;
$\ip$ is the pitch angle of the spirals at $R=R_a$.  A negative (positive)
value of $\ip$ indicates leading (trailing) spirals.

\begin{deluxetable}{lccccc}
\tabletypesize{\footnotesize}
\tablewidth{0pt}
\tablecaption{Properties of Nuclear Spirals\label{tbl:spiral}}
\tablehead{
\colhead{Model} &
\colhead{$R_{\rm in}$ (kpc)}  &
\colhead{$R_{\rm out}$ (kpc)}  &
\colhead{$\bSig/\Sigma_0$} &
\colhead{$\pSig/\bSig$} &
\colhead{$\ip$ (deg)}
}
\startdata
cs05bh0 &   0.13 &  0.50  &  0.41 &  7.92 &  $-33.8$ \\
cs05bh0t &   0.20 &  0.45  &  0.56 &  9.84 &  $-35.4$ \\
\hline
cs05bh7 &   0.02 &  0.45  &  0.92 &  1.95 & $8.5$    \\
cs10bh7 &   0.05 &  0.42  &  0.16 &  3.86 & $55.1$   \\
\hline
cs05bh8 &   0.02 &  0.55  &  1.00 &  2.07 & $3.5$    \\
cs10bh8 &   0.02 &  0.40  &  2.28 &  1.91 & $5.8$\\
cs15bh8 &   0.02 &$\cdots$&  8.09 &  1.67 & $6.2$\\
cs20bh8 &   0.02 &$\cdots$&  170.0&  1.74 & $7.8$
\tablecomments{
$R_{\rm in}$ and $R_{\rm out}$ are the inner and outer ends;
$\bSig$ and $\pSig$ are the mean and peak densities
at $R_a=0.05\kpc$ for Models cs15bh8 and cs20bh8
and $R_a=(R_{\rm in} R_{\rm out})^{1/2}$ for the other models;
$\ip$ is the pitch angle at $R=R_a$.
}
\end{deluxetable}

\begin{figure*}
\epsscale{1} \plotone{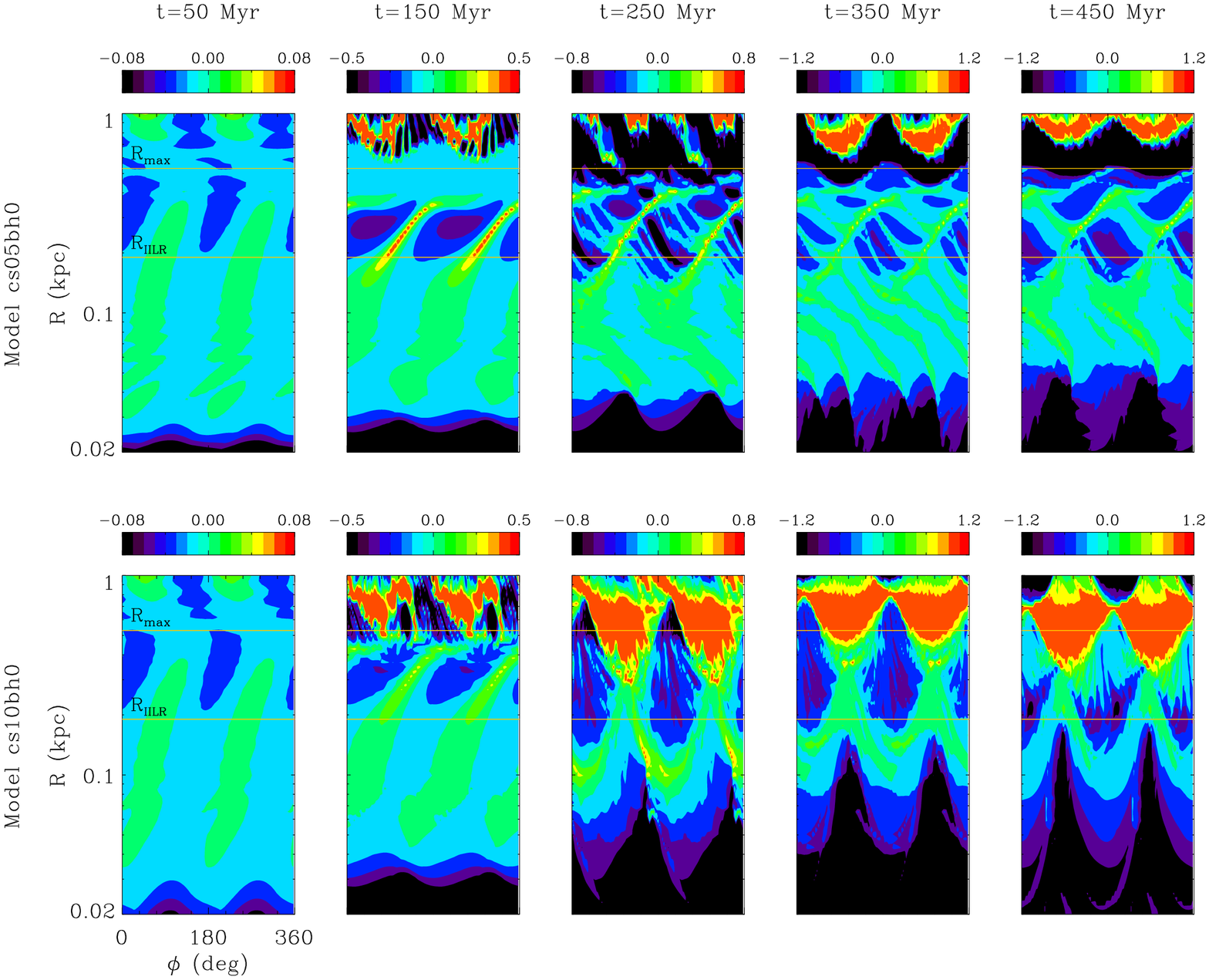}
\caption{Snapshots of
the logarithm of the gas
surface density on the $\phi-\log R$
plane for Models cs05bh0 (top row) and cs10bh0 (bottom row).
These
models do no have a central BH. Only
the regions with $R\leq1\kpc$ are shown. The locations of $\IILR$
and $\Rmax$ are indicated by two horizontal lines.
\label{fig:nspbh0}}
\end{figure*}

\subsection{Models Without A Black Hole}

To study the spiral features, it is convenient to show the logarithm
of the gas
surface density in logarithmic polar coordinates. Figure
\ref{fig:nspbh0} plots snapshots of gas surface density of Models
cs05bh0 and cs10bh0 on the $\phi-\log R$ plane. Any coherent
features with a positive (negative) slope on this plane are leading
(trailing) waves. Only the regions with $R\leq1\kpc$ are shown. Two
horizontal lines mark $\IILR$ and $\Rmax$ where the
$\Omega-\kappa/2$ curve is a locally maximum. At early time,  the
non-axisymmetric bar potential induces weak $m=2$ perturbations in
the central regions. Perturbed gas elements in the galactic plane
follow slightly elliptical orbits, which  are closed in a frame
rotating at $\Omega-\kappa/2$ and thus precess at a rate
$\Omega-\kappa/2$ near the ILRs when seen in a stationary bar frame. As
succinctly depicted in \citet{but96}, due to collisional dissipation
of gas kinetic energy occurring on converging orbits, the gas forms
spiral structures whose shape depends critically on the sign of
$d(\Omega-\kappa/2)/dR$ such that spirals are leading (trailing)
where $d(\Omega-\kappa/2)/dR$ is positive (negative). Figure
\ref{fig:nspbh0} indeed shows that when $t=50$ Myr the perturbed
density is leading at $R<\Rmax$ and trailing at $R>\Rmax$, although
the trailing features are soon overwhelmed by the nuclear ring.
Located away from the nuclear ring, however, the inner leading waves
are able to grow with time and eventually develop into shock waves
at $t\sim200$ Myr.  These nuclear spirals are short, extending over
$R\sim 0.13-0.50\kpc$, quite open with a pitch angle of
$\ip=-30^\circ$, and almost completely detached from the nuclear
ring.

\begin{figure}
\epsscale{1.1} \plotone{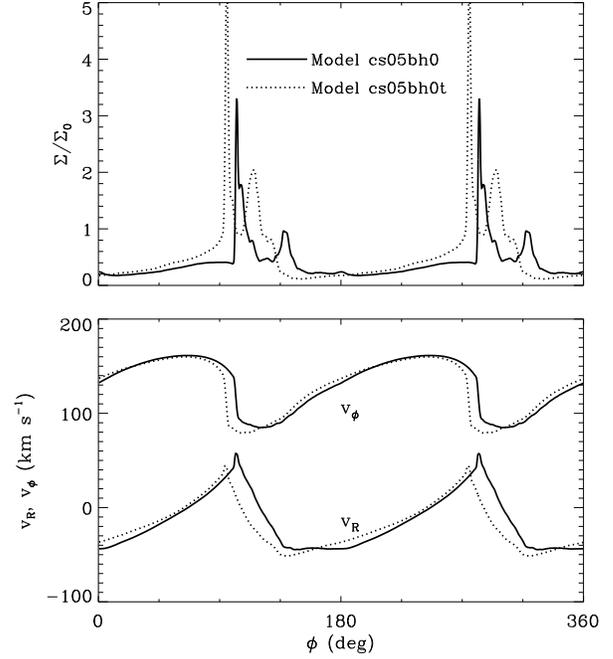}
\caption{
Azimuthal profiles of surface density (top) and velocities (bottom)
of the nuclear spirals at $R=0.25\kpc$ in Models
cs05bh0 (solid) and cs05bh0t (dotted) when $t=500$ Myr.
The spirals at this radius are shocks with
the compression factor of $\alpha\sim 3.4$.
\label{fig:nsprofbh0}}
\end{figure}

Figure \ref{fig:nsprofbh0} plots the azimuthal distributions of
surface density and velocities of the nuclear spirals at
$R=0.25\kpc$ in Models cs05bh0 (solid lines) and
cs05bh0t (dotted lines) when $t=500$ Myr.  Over the course of the orbits,
the changes of the radial and azimuthal velocities
associated with the spirals amount to $\sim 100\kms$,
which is indeed large enough to induce shocks.  The peak densities occurring
at $\phi\sim100^\circ, 280^\circ$ for Model cs05bh0  correspond to shock
fronts with a compression factor of $\alpha\sim3.4$. The density bumps at
$\phi\sim135^\circ, 315^\circ$ are produced by waves launched from the
inner boundary.
In Model cs05bh0t, the BH mass is increased to $\MBH\sim10^5\Msun$ due
to mass inflow, which supports slightly stronger, more leading spiral
shocks than in Model cs05bh0 with no BH.
Despite continual perturbations by traveling trailing waves
propagating from the inner boundary, the nuclear spirals in these models
last until the end of the run.
That leading nuclear spirals are persistent when the sound speed is
small is consistent with the results of SPH simulations reported by
\citet{ann05}.

The usual WKB dispersion relation for tightly-wound,
linear-amplitude waves in a non-self-gravitating medium reads
\begin{equation}\label{eq:disp1}
(\omega-m\Omega)^2 = k^2\cs^2 + \kappa^2
\end{equation}
where $\omega$ is the wave frequency and
$k$ and $m$ are the radial and azimuthal wavenumbers (e.g., \citealt{gol79}).
For $m=2$ waves corotating with a bar (i.e., $\omega=2\Omega_b$),
equation (\ref{eq:disp1}) becomes
\begin{equation}\label{eq:disp2}
k =  \pm \frac{2}{\cs} \sqrt{
(\Omega + \kappa/2 - \Omega_b)(\Omega-\kappa/2-\Omega_b)},
\end{equation}
(e.g., \citealt{eng00,mac04a}), indicating that nuclear spirals
corotating with the bar can exist only in the regions between
$\IILR$ and $\OILR$ when there are two ILRs\footnote{Fig.\
\ref{fig:nspbh0} shows that there are low-amplitude waves
propagating relative to the bar inside $\IILR$. Such waves can exist
inside the inner ILR as long as they satisfy equation
(\ref{eq:disp1}) in the WKB limit.}. However, the top row in Figure
\ref{fig:nspbh0} reveals that the nuclear spirals in Model cs05bh0
extend slightly inward of $\IILR$. This seemingly contradicting
result is due to nonlinear effects which are not captured by WKB
theory. The velocity perturbations associated with these spirals are
so large that fluid elements just outside $\IILR$ can move across
the inner ILR over the course of their epicycle orbits, providing
perturbations for the gas at $R<\IILR$ that responds passively. In
Model cs05bh0, the radial velocity perturbation is $\Delta
v_R=52\kms$. Since the epicycle frequency is $\kappa=1100\freq$ at
$R=\IILR$,  the corresponding radial amplitude of the epicycle
orbits is estimated to be $\Delta R=\Delta v_R/\kappa =0.05\kpc$,
which is in good agreement with the radial extent of the nuclear
spirals inward of $\IILR$.

Models without a BH and with $\cs\geq 10\kms$ do form nuclear spirals at
early time,
but they are all transient, lasting less than 200 Myr.
The bottom row of Figure \ref{fig:ringall} shows how nuclear spirals
are destroyed in Model cs10bh0.
The nuclear ring in this model is not only located inside $\Rmax$ but also
has large thermal pressure, continuously generating sonic perturbations
that propagate radially inward.
Because of the background shear, the perturbations are preferentially in
the form of trailing waves which interact destructively with the leading
spirals that formed at $R<\Rmax$, destroying the latter.
The destruction of nuclear spirals happens at $t=185$ Myr for Model
cs10bh0.  This occurs earlier as $\cs$ increases, since disturbing pressure
perturbations are correspondingly stronger.

\begin{figure*}
\epsscale{1} \plotone{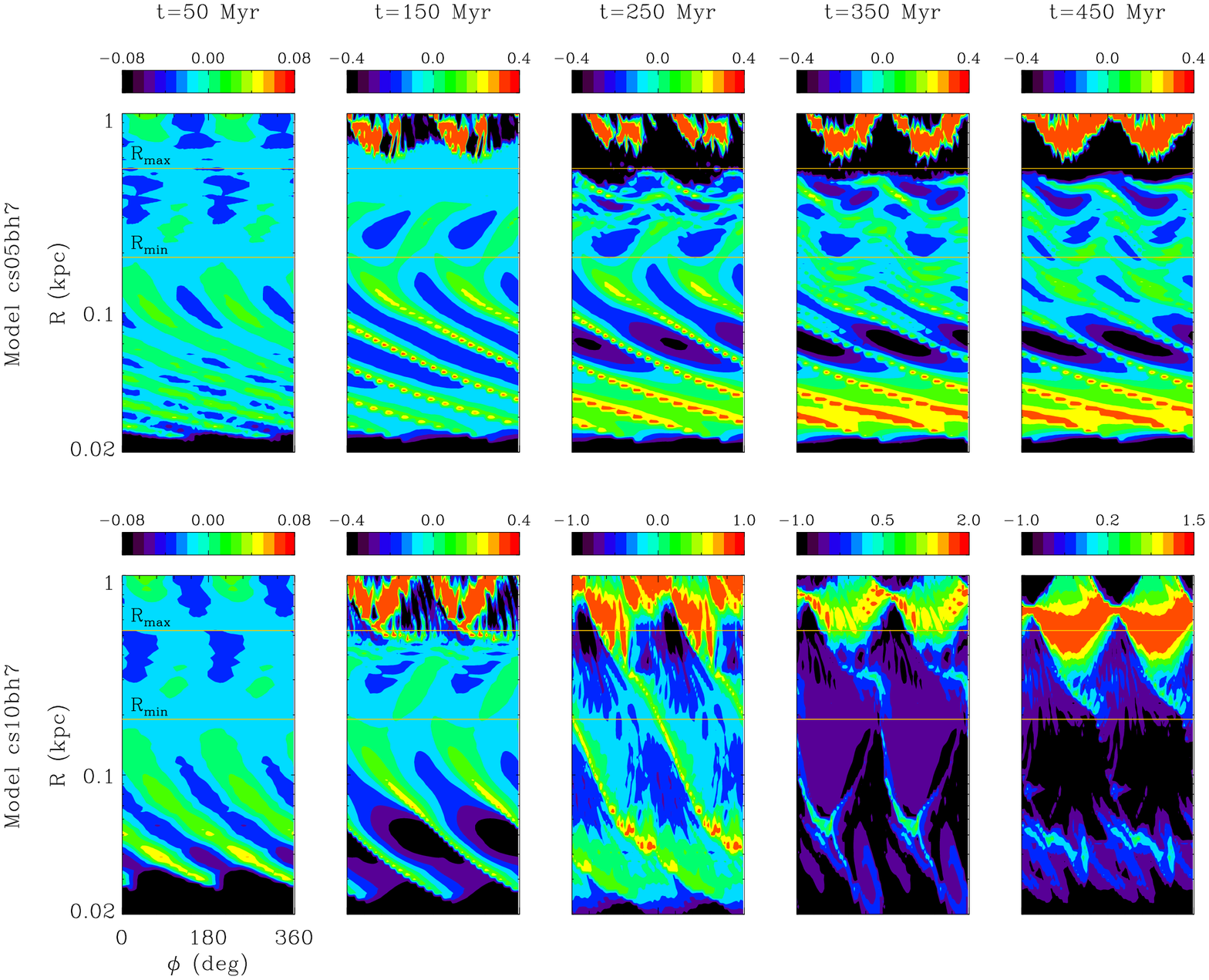}
\caption{Snapshots of the logarithm of the gas
surface density on the $\phi-\log R$
plane for Models cs05bh7 (top row) and cs10bh7 (bottom row).  These
models have a central BH with a mass of $4\times10^7\Msun$. Only
the regions with $R\leq1\kpc$ are shown. The locations of $\Rmax$
and $\Rmin$ are indicated by two horizontal lines.
\label{fig:nspbh7}}
\end{figure*}

\subsection{Models With $\MBH=4\times10^7\Msun$}

Since the presence of a central BH greatly changes
the $\Omega-\kappa/2$ curve in the central regions, it is interesting
to explore how the morphologies of nuclear spirals depend on the
BH mass.   In bh7 and bh8 models with a single ILR, stationary waves
in the bar frame can exist inside $\ILR$ all the way down to the center.
Figure \ref{fig:nspbh7} plots snapshots of gas surface density
for Models cs05bh7 and cs10bh7 on the logarithmic polar plane.
The positions of $\Rmax$ and $\Rmin$ are indicated by the horizontal lines.
As expected, the overdense perturbations produced by orbit crowding
at early time ($t=50$ Myr) have leading configurations at $\Rmin<R<\Rmax$
and trailing configurations at $R<\Rmin$ or $R>\Rmax$.
The overdense regions at $R>\Rmax$ are subsequently wiped out as the
nuclear ring forms,
while those at $R<\Rmax$ grow into trailing nuclear spirals.
In Model cs05bh7,
the leading spirals at $\Rmin<R<\Rmax$ also grow slightly until $t\sim150$
Myr to temporarily form ``inner-trailing and outer-leading'' structures
represented by the double cross-hatching in Figure \ref{fig:nsp_class}.
As trailing perturbations from both the inner trailing spirals and the outer
nuclear ring propagate and interfere with the leading spirals,
it becomes increasingly difficult to identify coherent spiral structures
at $\Rmin<R<\Rmax$.  On the other hand, the trailing
spirals at $R<\Rmin$ keep growing until $t\sim230$ Myr after which their
amplitude of $\pSig/\bSig\approx2.0$ remains more or less constant.
They are approximately logarithmic in shape with a pitch angle of
$\ip=8.5^\circ$.
Unlike in Model cs05bh0 where leading spirals are actually shocks,
the trailing nuclear spirals in Models cs05bh7 are relatively weak
(with the maximum compression factor of $\alpha_{\rm max}\sim0.28$
at $t=500$ Myr),
and never develop into shocks.

In Model cs10bh7 (bottom row of Fig.\ \ref{fig:nspbh7}), the nuclear
spirals have larger $k$ and thus are more open than those in the
$\cs=5\kms$ counterparts (see, e.g., eq.\ [\ref{eq:disp2}]).  Since
the nuclear ring in this model forms at $\Rring\approx0.5\kpc$, it
can directly destroy the leading spiral at $\Rmin<R<\Rmax$, and
feeds the inner trailing spirals by supplying trailing
perturbations.  Thus, the inner spirals grow both in strength and
spatial extent to make contact with the nuclear ring at $t=210$
Myr. At this time, all parts of the nuclear spirals turn to shocks
with the maximum density of $\Sigma_{\rm max}/\Sigma_0=3.8$
and the corresponding compression factor of $\alpha=0.8$
at $R=0.1\kpc$. Gas passing through the spiral shocks loses angular
momentum, increasing the mass inflow rate at the inner boundary. As
the amount of gas lost in the nuclear regions increases, the density
of the nuclear spirals decreases with time, but they remain as
shocks with the compression factor of $\alpha\sim1.3$
until the end of the run.

In Models cs15bh7 and cs20bh7, nuclear spirals start out with inner-trailing
and outer-leading shapes, and evolve into trailing-only configurations, as in the
other bh7 models.  However, the highly eccentric orbits of the gas
affected by thermal pressure dismantle the inner spiral structures almost
completely in these models: thermal perturbations are so strong that
a central BH with $\MBH=4\times10^7\Msun$ cannot enforce circular orbits
in the very nuclear regions (see Fig.\ \ref{fig:st_ring}c).

\subsection{Models With $\MBH=4\times10^8\Msun$}

\begin{figure*}
\epsscale{1} \plotone{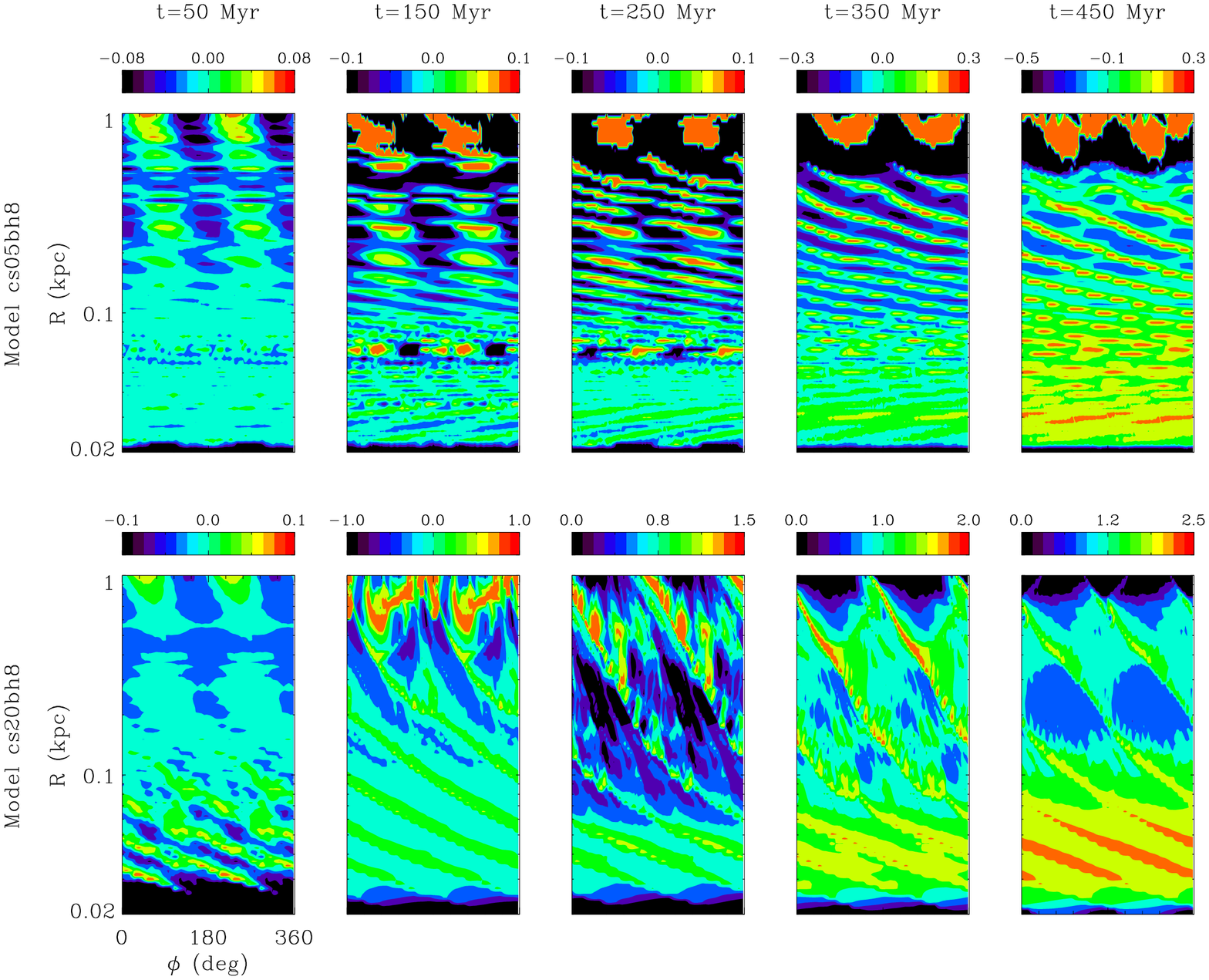}
\caption{ Snapshots of
the logarithm of the gas
surface density on the $\phi-\log R$
plane for Models cs05bh8 (top row) and cs20bh8 (bottom row).  These
models have a central BH with a mass of $4\times10^8\Msun$.  Only
the regions with $R\leq1\kpc$ are shown. \label{fig:nspbh8}}
\end{figure*}

Since the $\Omega-\kappa/2$ curve decreases monotonically with $R<1\kpc$
in bh8 models, nuclear spirals, if they exist, are all trailing as evident in
Figure \ref{fig:nspbh8}.
For Model cs05bh8 (top row of Fig.\ \ref{fig:nspbh8}), the spirals
evolve almost independently of, and are well separated from,
the nuclear ring.
At $t=500$ Myr, they have a very small pitch angle $\ip=3.5^\circ$
corresponding to $kR=2\cot \ip=33$ owing to the strong background
shear.  They are also very weak, with an amplitude of
$\pSig/\bSig=2.1$ at $R\approx0.1\kpc$.
When $\cs=10\kms$, the pitch angle is increased to $\ip=5.8^\circ$
and the nuclear spirals extend outward all the way to the nuclear ring.
Except near the contact points, the spirals are still decoupled from
the ring.  With quite a large value of $kR=20$, the component of the
rotational velocity perpendicular to the spirals is not large enough
to induce shocks.

We have seen earlier that the large thermal pressure in the rings of
models with $\cs\geq15\kms$ provides strong perturbations that make
the gas orbits near the galaxy center highly eccentric, destroying
nuclear spirals in bh0 and bh7 models.  In bh8 models, however, the
situation is quite different since a central BH with
$\MBH=4\times10^8\Msun$ dominates the potential, keeping the orbits
almost circular in the very central regions. Even with a large thermal
pressure, the gas orbits there cannot be very eccentric, so that
the nuclear spirals are protected from disruptive pressure
perturbations. On the other hand, the ring material is quite
distributed because of the large pressure gradients,
feeding a trailing nuclear
spiral that grows strongly in Models cs15bh8 and cs20bh8.

\begin{figure}
\epsscale{1.1} \plotone{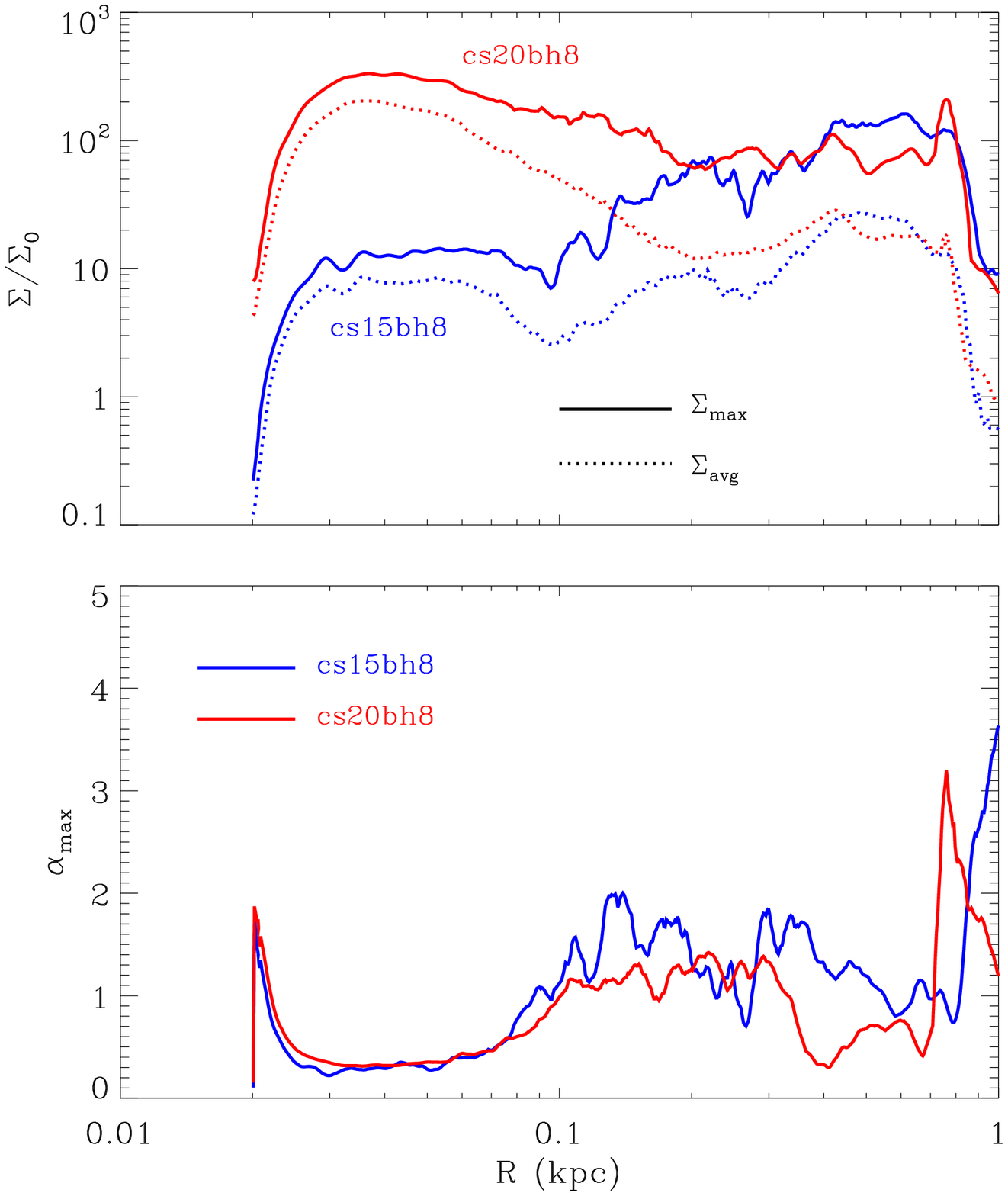}
\caption{Radial distributions of the maximum and mean densities (top)
and the maximum compression factor (bottom) in the inner $1\kpc$ regions of
Models cs15bh8 and cs20bh8 at $t=500$ Myr.
Due to strong spiral shocks, the ring material in Model cs20bh8 is already
moved to the $R\sim0.02-0.1\kpc$ region, while with weaker shocks it still
lies at $R>0.1\kpc$ in Model cs15bh0.
In both models, the shocks at $R\sim0.05\kpc$ are quite weak with
the compression factor of $\alpha\sim0.3$.
\label{fig:nsprofbh8}}
\end{figure}

At $t=120$ Myr, the spirals in Models cs15bh8 and cs20bh8 turn into shocks
and touch the densest parts of the ring located at $R\sim0.4\kpc$.
Because of the larger pitch angles, the shocks are stronger in the outer parts;
the portions at $R\simlt0.1\kpc$ are weak shocks with
density jumps of only $\sim 2$.
Similarly, the shocks in Model cs20bh8 are stronger than in Model cs15bh8
since the former has more open spirals and a denser ring.
In both models,
the shocks near the ring are so strong that even the gas constituting
the ring suffers from a significant loss of angular momentum at the
intersecting points.  At $t=150$ Myr, the ring material is essentially
dissected by the trailing spiral shocks and gradually moves toward the center.
As the ring material continues to flow in, the spirals appear as a direct
continuation of the off-axis shocks ($t=350$ Myr), consistent with
the results of \citet{mac04b}.

Figure \ref{fig:nsprofbh8} plots the radial distributions of the maximum
and mean densities as well as the maximum compression
factor in the inner $1\kpc$ regions of Models cs15bh8 and cs20bh8 at
$t=500$ Myr.
The inflow of the ring material in Model cs20bh8 is quite strong
that the gas is collected in the nuclear regions with $R\sim 0.02-0.1\kpc$;
the amount of the gas that goes out through the inner boundary
is much smaller than that comes in.
With weaker shocks and thus less angular momentum loss,
on the other hand, the destroyed ring gas in Model cs15bh8 is
still mostly at $R>0.1\kpc$ at this time.
Note that the density enhancement at $R=0.05\kpc$ resulting from the
dissolution of the ring is $\sim8\Sigma_0$ and $\sim170\Sigma_0$
in Models cs15bh8 and cs20bh8, respectively:
the mean contrast of the spirals that are weak shocks with
the compression factor $\alpha\sim 0.3$
is $\pSig/\bSig\sim1.6-1.8$ at $R=0.05\kpc$ in both models.
This increase of the gas surface density in the central regions
is the primary reason for enhanced mass inflow rates at late time
in bh8 models with large $\cs$.

\section{Mass Inflow Rates}

Galactic bars are considered to be a promising means of transporting gas to the
centers of galaxies to fuel supermassive BHs and produce AGNs.  Since our
numerical models use a cylindrical grid with a circular boundary,
they are ideally suited to study how the mass accretion rate depends on
the gas sound speed and the BH mass.   We assume that all the gas that
crosses the inner boundary located at $R=20\pc$ in our models is accreted
to the central BH.
In reality, a large amount of mass in gas near the BH may change the gas
orbits by providing pressure and gravitational forces.
Therefore, $\Mdot$ through the inner boundary that we measure
is likely to be an upper limit to the real mass accretion rate to the BH.
In addition, $\Mdot$ is likely to depend on the inner
boundary size especially when gas orbits are highly eccentric near the
inner boundary.
Note that in non-self-gravitating, isothermal, and unmagnetized systems,
$\Mdot$ resulting from simulations is linearly proportional to the
adopted initial surface density $\Sigma_0$;
all the models presented here take $\Sigma_0=10\Surf$.

\begin{deluxetable}{lcc}
\tabletypesize{\footnotesize}
\tablewidth{0pt}
\tablecaption{Total Mass of Gas Inflow at $t=500$ Myr\label{tbl:macc}}
\tablehead{
\colhead{Model} &
\colhead{$\Macc (\Msun)$}
}
\startdata
cs05bh0 &  $8.9\times10^4$  \\
cs05bh0t & $1.1\times10^5$   \\
cs10bh0 &  $8.0\times10^5$ \\
cs15bh0 &  $2.7\times10^6$ \\
cs20bh0 &  $1.8\times10^7$ \\
cs20bh0t & $2.3\times10^7$  \\
\hline
cs05bh7 &  $4.2\times10^4$  \\
cs10bh7 &  $1.6\times10^6$ \\
cs15bh7 &  $7.4\times10^6$ \\
cs20bh7 &  $3.2\times10^7$ \\
cs20bh7t & $2.9\times10^7$ \\
\hline
cs05bh8 &  $9.4\times10^3$ \\
cs10bh8 &  $4.5\times10^4$ \\
cs15bh8 &  $2.0\times10^5$ \\
cs20bh8 &  $5.1\times10^6$
\end{deluxetable}

\begin{figure}
\epsscale{1.2}\plotone{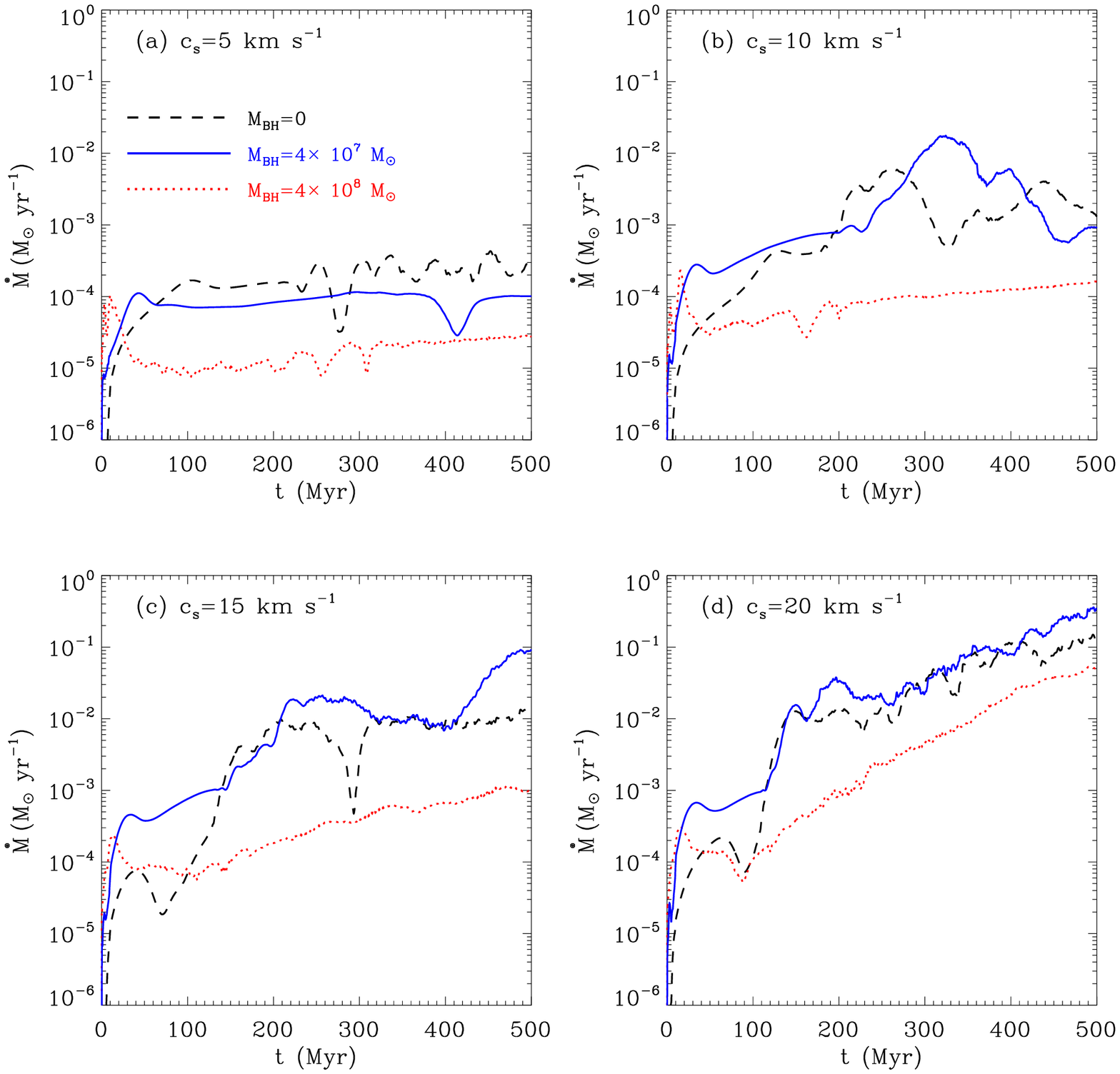}
\caption{
Temporal evolution of the mass inflow rates $\Mdot$ through the inner boundary
at $R=20\pc$ for models with
(a) $\cs=5\kms$, (b) $\cs=10\kms$, (c) $\cs=15\kms$, and (d) $\cs=20\kms$.
The dashed, solid, and dotted lines correspond to the cases with
$\MBH=0$, $4\times10^7\Msun$, and $4\times10^8\Msun$, respectively.
\label{fig:mdot}}
\end{figure}

In general, $\Mdot$ would be large if the gas orbits near the center
are highly eccentric or radial, while circular orbits
would make $\Mdot$ quite small.  This expectation is consistent with
Figure \ref{fig:mdot} which plots the temporal evolution of the mass
inflow rates for all models.
The corresponding accreted gas mass $\Macc = \int \Mdot(t)dt$ over
$500$ Myr is given in Table \ref{tbl:macc}.
Clearly, $\Mdot$ is larger for models with larger $\cs$ and
no BH, compared to models with a massive BH of $\MBH=4\times10^8\Msun$.
When $\cs=5\kms$, the nuclear spirals are
well separated from the nuclear rings, and the departure of the gas orbits
from a circular shape near the inner boundary is small, resulting
in quite small values of $\Mdot$ ($\simlt 10^{-3}\Aunit$). In bh8 models,
the presence of a central BH makes $\Mdot$ smaller by about an
order of magnitude than in bh0 models by providing strong axisymmetric
gravitational potential near the center.  In bh7 models, the BH potential
is not strong enough to circularize the eccentric orbits,
giving rise to $\Mdot$ only slightly smaller than that in bh0 models.

Increasing $\cs$ enhances $\Mdot$ because pressure perturbations become
stronger and the nuclear ring tends to be located
closer to the center, both of
which strongly affect the gas orbits in the central region.
When $\cs=10\kms$, $\Mdot$ is increased by about an order of magnitude
compared to the cases with $\cs=5\kms$, but is still less than
$10^{-2}\Aunit$ except for a brief time interval around
$t=300$ Myr in Model cs100bh7 when the nuclear spirals shock the
central gas and cause enhanced inflows.  When $\cs\geq 15\kms$,
the gas orbits in bh0 and bh7 models are quite eccentric near the center,
so that some gas with highly radial orbits plunges directly
into the inner boundary, increasing $\Mdot$ dramatically
compared to the lower $\cs$ counterparts.
The associated gas mass accreted to the galaxy center is of order
of $\sim10^7\Msun$ over 500 Myr, suggesting that a strong galactic
bar like studied in this work can be an appealing means for the
growth of supermassive BHs provided the gaseous medium has
a large (effective) sound speed (cf.~\citealt{shl89}; see also,
e.g., \citealt{vol10} for review).
In bh8 models, on the other hand, the gas orbits in the
central parts are not perturbed much (since the initial angular momentum is
overwhelmingly large). As the nuclear spirals grow into shocks, however,
the nuclear gas as well as the ring material in these models drift inward
slowly, increasing $\Mdot$ over time.
In models with $\MBH=4\times10^8\Msun$,
the late-time values of $\Mdot$ in models with $\cs=15$ or $20\kms$ are
larger than $0.01\Aunit$, sufficient to power AGNs in Seyfert galaxies
(e.g., \citealt{fri93,fab08}).

\begin{figure}
\epsscale{1.2}\plotone{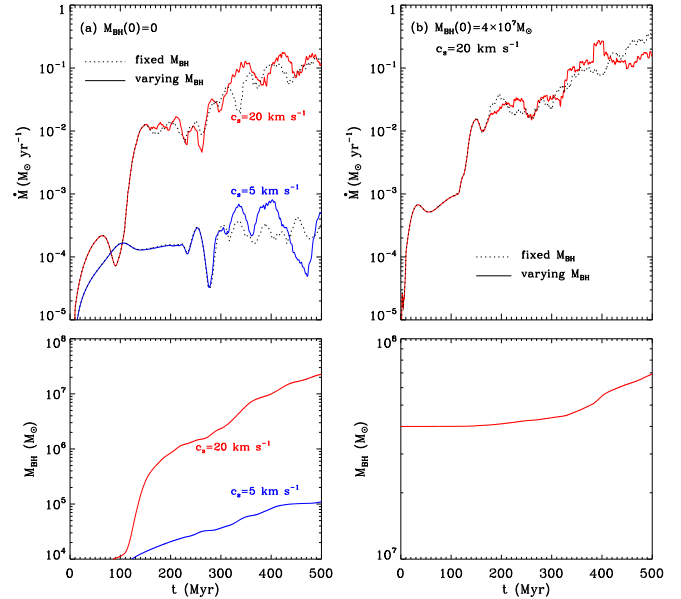}
\caption{
Temporal evolution of the mass inflow rates (top) and the BH
mass (bottom) in Models cs05bh0t, cs20bh0t, and cs20bh7t where
$\MBH$ is allowed to vary with time.
The mass inflow rates from the fixed-$\MBH$ counterparts are compared as
dotted lines.  In all models, the total increase in the BH mass over
500 Myr is not large enough to cause significant changes in $\Mdot$.
\label{fig:mdot_MBHt}}
\end{figure}

Finally, we present the mass inflow rates resulting from the models in which
$\MBH$ is varied self-consistently with $\Mdot$.
The left panels of Figure \ref{fig:mdot_MBHt}
compare $\Mdot$ from Models cs05bh0t and cs20bh0t with those from the
fixed-$\MBH$ counterparts, while the bottom panels plot the
temporal evolution of the BH mass in the former models.
In Model cs05bh0t, the total gas mass accreted over 500 Myr is
$\sim 10^5\Msun$, with the corresponding increment of the equilibrium
rotational velocity of $\sim4\kms$ at the inner boundary.  Since this is
three times smaller than the initial circular velocity there, the
BH mass does not affect $\Mdot$ much,
as Figure \ref{fig:mdot_MBHt} illustrates.
In the case of Model cs20bh0t, $\MBH$ attains $\sim2\times10^7\Msun$ at
$t=500$ Myr, which is large enough to be dynamically important.
When $\cs=20\kms$, however, the gas orbits are highly eccentric
and $\Mdot$ is insensitive to the BH mass as long
as $\MBH\simlt4\times10^7\Msun$ (see Fig. \ref{fig:mdot}d).
As the rignt panels of Figure \ref{fig:mdot_MBHt} show,
the increase of $\MBH$ over 500 Myr in Model cs20bh7t is
less than a factor of two, corresponding to the equilibrium circular
velocity 1.3 times the initial value.  With an enhanced
centrifugal barrier, the resulting $\Mdot$ becomes gradually smaller
than the case with fixed $\MBH$, but by less than, on average,
$\sim10\%$ in $t=300$-$500$ Myr.
Therefore, we conclude that the effect of BH growth due to gas
accretion on the mass inflow rate as well as bar substructures
is not significant for the models we have considered.

\section{Summary and Discussion}

\subsection{Summary}

We have presented detailed numerical models that explore the formation of
substructures produced by the gas flow in barred galaxies.
Previous models based on particle simulations
(e.g., \citealt{eng97,ann05,tha09}) did not have
sufficient resolution to resolve nuclear spirals.  On the other hand,
studies that used
the grid-based code CMHOG (e.g., \citealt{pin95,mac04b}) unknowingly made
mistakes in the force evaluation for the bar potential, so that
the results needed to be recomputed.

In this paper we have corrected the errors in the original CMHOG code and
run high-resolution hydrodynamical simulations. To resolve the
nuclear regions, we employ a logarithmically-spaced cylindrical
grid, with a zone size of $\Delta R \leq 6\pc$ at $R\leq1\kpc$ where
nuclear rings and spirals form.  We have included the potential
from a central BH, and studied the flow properties as the mass of the BH
$\MBH$ and the sound speed $\cs$ in the gas are varied.
For simplicity, the effects of gaseous self-gravity and magnetic fields
are not included.
The main results of the  present work are summarized as follows:

1. \emph{Off-axis Shocks} -- The imposed non-axisymmetric bar
potential provides gravitational torques to the otherwise
circular-rotating gas, perturbing its orbit.  The perturbed orbits
crowd at the downstream sides of the bar major axis and produce
overdense ridges that eventually develop into off-axis shocks. At a
quasi-steady state, the off-axis shocks are overall almost parallel
to $\xone$-orbits:  they start from the bar major axis at the outer
ends, are gradually displaced downstream as they move inward,
and connect to the nuclear ring at the inner ends. While the
positions of the off-axis shocks are almost independent of $\MBH$,
since the effect of a BH is negligible at large radii, they depend
on $\cs$ in such a way that the shocks are, on average, located
closer to the bar major axis as $\cs$ increases.  This is primarily
because gas with larger $\cs$ should be more strongly perturbed to
induce shocks, which occurs deeper in the potential well and thus
results in shocks on lower $\xone$-orbits \citep{eng97}.

The off-axis shocks are in general curved. Flow streamlines are
complicated near the shocks in that they diverge before the shocks
and are promptly swept inwards by inflowing gas right after the
shocks.  Therefore, the usual Ranking-Hugoniot jump conditions for
planar, one-dimensional shocks are not applicable to the off-axis
shocks. In fact, the shock strength, as measured by the peak density
$\Sigma_{\rm max}$ at $R\sim(1.5-1.8)\kpc$ from the center, is
$\Sigma_{\rm max}/\Sigma_0\sim3-6$ for models with no BH and do not
sensitively depend on $\cs$.
The compression factor of the off-axis shocks are insensitive to
the BH mass and depends on $\cs$
roughly as $\alpha_{\rm max}\sim 7.7 (\cs/5\kms)^{0.92}$.
The off-axis shocks have very strong
velocity shear amounting to $\sim(1-3)\times10^3\kms\kpc^{-1}$. This
strong shear may suppress the growth of gravitational instability in
the high-density off-axis shocks when self-gravity is included.

2. \emph{Nuclear Rings} -- When gas passes through the off-axis
shocks, it loses angular momentum, flows inward, and forms a nuclear
ring where the centrifugal force balances the external gravity. The
ring is attached to the inner ends of the off-axis shocks and thus
becomes smaller in size as $\cs$ increases. The rings that form in
our models are all located inside the (outer) ILR, but this does not
imply that the ring formation is related to the ILRs. When $\cs$ is
small, the pressure perturbations on the gas orbits are so weak that
the rings are quite narrow and their shape is well described by
$\xtwo$-orbits. The  mean radius is $\Rring\sim(0.8-0.9)\kpc$ when
$\cs=5\kms$ and $\Rring\sim(0.5-0.6)\kpc$ when $\cs=10\kms$,
independent of the BH mass. This suggests that the ring position is
not determined by the $\Omega-\kappa/2$ curve, hence by the ILRs.
When $\cs \geq 15\kms$, on the other hand, large thermal pressure
strongly affects the gas orbits in the nuclear rings. For example,
some gas near the contact points between the ring and the off-axis
shocks is forced out to follow relatively round orbits, while other
gas is pulled in radially to make very eccentric orbits.  These
diverse gas orbits near the contact points tend to spread out the
ring material, making the rings much broader than in models with
smaller $\cs$.

3. \emph{Nuclear Spirals} -- Since even the non-axisymmetric bar
potential is nearly axisymmetric in the central parts, the gaseous
responses inside a nuclear ring are not as dramatic as in off-axis
shocks.  Nevertheless, non-axisymmetric $m=2$ perturbations are able
to grow inside a ring and develop into nuclear spirals that persist
for a long period of time, provided that $\cs$ is small or $\MBH$ is
large. Although all models have weak spiral structures at early
time, in models with large $\cs$ they are soon destroyed by
eccentric gas orbits as well as perturbations induced by the large
pressure in the rings unless the BH mass is very large. When
$\MBH=4\times10^8\Msun$,  the disruptive pressure perturbations from
the rings cannot penetrate the very central parts where the gas has
extremely large initial angular momentum.  In this case, the nuclear
spirals are protected from the surrounding and thus are long-lived.

The shape of nuclear spirals is determined by the sign of
$d(\Omega-\kappa/2)/dR$ such that spirals that form in the regions
where $d(\Omega-\kappa/2)/dR$ is positive (negative) are leading
(trailing), confirming the theoretical expectations of
\citet{but96}. With no BH, only the model with $\cs=5\kms$ has
persistent leading spirals in the regions where the
$\Omega-\kappa/2$ curve is an increasing function of $R$. The
leading spirals in this model are quite strong and develop into
shocks with the peak density $\pSig/\bSig\sim7.9$ and the
compression factor $\alpha\sim 3.4$ at $R=0.25\kpc$ at the end of
the run. Models with $\MBH=4\times10^7\Msun$ initially have hybrid
features comprising of trailing spirals at $R<\Rmin$ and leading
spirals at $\Rmin<R<\Rmax$, where $\Rmin$ and $\Rmax$ refer to the
radii where $\Omega-\kappa/2$ curve attains a local minimum and
maximum, respectively.  When $\cs\leq10\kms$, however, the leading
parts in the hybrid spirals are destroyed by the trailing pressure
waves launched by the ring, leaving only the weak trailing spirals
behind. When $\cs\geq15\kms$, both leading and trailing spirals are
destructed completely by the pressure perturbations. In models with
$\MBH=4\times10^8\Msun$, $d(\Omega-\kappa/2)/dR<0$ in the whole
nuclear regions, so that nuclear rings are always trailing. When
$\cs\leq10\kms$, the spirals well separated from the ring are weak
and tightly wound. When $\cs\geq15\kms$, on the other hand, the
trailing spirals are fed by the gas in the nuclear ring and grow
into shocks, the outer ends of which join the inner ends of the
off-axis shocks smoothly. Although the nuclear spirals in these
models have only modest density contrasts of $\pSig/\bSig\sim1.7$ at
$R=0.05\kpc$, the background density resulting from the inflows of
the ring material is greatly enhanced (by more than two orders of
magnitude when $\cs=20\kms$).

4. \emph{Mass Inflow Rates} --
Although gas experiences a significant loss in angular momentum  when
it meets off-axis shocks during galaxy rotation, this does not necessarily
translate into the mass inflow all the way to the galaxy center.
In models with $\cs\leq10\kms$, a narrow nuclear ring formed by gas
with $\xtwo$-orbits inhibits further inflows of the gas, resulting
in the mass inflow rate $\Mdot$ through the inner boundary less than
$0.01\Aunit$, regardless of the BH mass.  The mass inflow rates are
greatly enhanced to $\Mdot > 0.01\Aunit$ in models with $\MBH\leq
4\times10^7\Msun$ and $\cs\geq15\kms$ or
$\MBH=4\times10^8\Msun$  and $\cs=20\kms$
for different reasons depending on the BH mass.
When $\MBH\leq 4\times10^7\Msun$, the gas orbits are affected by
thermal pressure and some gas in the ring can take highly eccentric
orbits, directly falling into the inner boundary.  In models with
$\MBH= 4\times10^8\Msun$, on the other hand, the gas orbits near the
center are more or less circular, but the density in the nuclear
regions is enhanced greatly because of strong nuclear spirals,
increasing $\Mdot$.

\subsection{Discussion}

Nuclear rings play an important role in evolution of barred galaxies
by providing sites of active star formation near the centers (e.g.,
\citealt{but86,gar91,bar95,mao01,maz08}).  Rings certainly consist
of gas that migrates from outer parts inward by losing angular
momentum at off-axis shocks, but what stops further migration to
form a ring remains a matter of debate. As a trapping mechanism of
the ring material, \citet{com96} proposed the non-axisymmetric bar
toque that forces gas to accumulate between two ILRs or at a single
ILR depending on the shape of the gravitational potential (see also
\citealt{but96}), while \citet{reg03,reg04} favored the gas
transitions from $\xone$- to $\xtwo$-orbits rather than orbital
resonances.  However, our numerical results show that a ring is
formed at early time because of the centrifugal barrier that the
migrating material feels. Later on, nuclear rings slowly shrink in
size as gas with lower angular momentum gas is continuously added.
Although nuclear rings are located in between two ILRs in models
with no BH, this is a coincidence.  Models with a central BH show
that the specific ring positions are insensitive to the shape of the
$\Omega-\kappa/2$ curves, suggesting that nuclear ring formation is
not a consequence of the orbital resonances. In fact, the gas flows
that produce the ring are not in force balance and have a large
radial velocity, so that the concept of resonances and ILRs is not
applicable to nuclear rings (e.g., \citealt{reg03}). In addition,
the notion of $\xtwo$-orbits as the gas trapping locations is
meaningful only for small $\cs$.\footnote{The models considered by
\citet{reg03,reg04} had the sound speed fixed to $\cs=5\kms$.} When
the sound speed is large, thermal pressure at the contact points
between the off-axis shocks and nuclear ring causes the gas orbits
to deviate from $\xtwo$-orbits considerably.

The results of our simulations suggest that not all barred-galaxies
possess nuclear spirals at their centers.  Long-lasing nuclear
spirals exist only when either the sound speed is small or the BH
mass is large: they do not survive in models with \emph{both} large
$\cs$ \emph{and} small $\MBH$. Two common views regarding the nature
of nuclear spirals are low-amplitude density waves and strong
gaseous shocks (see e.g.,
\citealt{eng00,mac02,mac04a,mac04b,ann05,tha09}).
And, our simulations indeed show that they are either tightly-wound
density waves or shocks when the pith angle is relatively large
($\ip>6^\circ$).  One may  speculate
that nuclear spirals are more likely to be shocks rather than
density waves when $\cs$ is small. In contrast to this prediction,
however, nuclear spirals in models with $\MBH=4\times10^8\Msun$ are
density waves when $\cs$ is small and shocks when $\cs$ is large.
This is of course because as $\cs$ increases, waves in nuclear
regions tend to be more open (with smaller $|k|$) in the beginning,
and they are subsequently supplied with more gas from the rings as
they grow.
This is entirely consistent with the results of \citet{ann05} who used
SPH simulations to show that nuclear spirals are supported by
shocks when $\cs\simgt15\kms$ in models with a massive BH.
We note however that weak trailing spirals seen
in our Model cs10bh8 are absent in Model M2 ($\cs=10\kms$ and
$\MBH=4\times10^8\Msun$) of \citet{ann05}, which is presumably due to
an insufficient number of particles to resolve nuclear spirals in their SPH
simulations.

Of 12 models with differing $\cs$ and $\MBH(0)$ that we have
considered, only 1 model possesses leading spirals, suggesting that
galaxies with leading nuclear spirals would be very uncommon in
nature.  To our knowledge, only two galaxies, NGC 1241 and NGC 6902,
are known to possess leading features in the nuclear regions
\citep{dia03,gro03}\footnote{Leading arms in NGC 1241 are detected
by Pa$\alpha$ emissions tracing young stars, while those in NGC 6902
are observed in the K$^\prime$ band tracing old populations. It is
uncertain how these stellar features are related to gaseous nuclear
spirals studied in this paper.}. Based on our simulations, the
existence of leading spirals at centers requires two stringent
conditions: (1) the gas should be dynamically cold enough to protect
nuclear spirals from nuclear rings and (2) there should be a wide
range of radii with $d(\Omega-\kappa/2)/dR<0$ in the central parts,
which can be easily accomplished when there are two ILRs (or without
a strong central mass concentration). The second condition is
consistent with the linear theory that predicts short leading waves
propagating outward from the inner ILR \citep{mac04a}. The facts
that NGC 6902 is a barred-spiral galaxy \citep{lau04} and does not
show significant $X$-ray emissions indicative of AGN activities
\citep{des09} are not inconsistent with the second requirement for
the existence of leading nuclear spirals. Since NGC 1241 is a
Seyfert 2 galaxy with an estimated BH mass of $\log
(\MBH/\Msun)=7.46$ \citep{bia07}, however, the nuclear star-forming
regions in this galaxy are unlikely to be associated with gaseous
nuclear spirals studied in this work.

Finally, we discuss the mass inflow rates derived in our models in
regard to powering AGNs in Seyfert galaxies.   The mass accretion
rate is often measured by the Eddington ratio defined by $\lambda
\equiv L_{\rm bol}/L_{\rm Edd} = 4.5\times10^{-2} (\epsilon/0.1)
(\Mdot/10^{-2}\Aunit)(\MBH/10^7\Msun)^{-1}$, where $L_{\rm bol}$ and
$L_{\rm Edd}$ denote the bolometric and Eddington luminosities of an
AGN and $\epsilon$ is the mass-to-energy conversion efficiency of
the accreted material. Observations indicate that $\lambda\simlt
0.1$ for classical Seyfert 1 galaxies with broad iron emission lines
(e.g., \citealt{mey11}; see also \citealt{pet97}), while
$\lambda\sim 10^{-3}$ for low-luminosity Seyfert 1 AGNs (e.g.,
\citealt{ho08}). In our numerical models, the mass inflow rates are
larger for models with smaller $\MBH$ and larger $\cs$. Taking
$\epsilon\approx0.1$ (e.g., \citealt{yu02}), the mass inflow rates
amount to $\lambda \simlt 10^{-4}$ for $\MBH=4\times10^8\Msun$ and
$\cs\simlt15\kms$, $\lambda \sim 10^{-3}$ for
$\MBH=4\times10^8\Msun$ and $\cs=20\kms$ or for
$\MBH=4\times10^7\Msun$ and $\cs=10\kms$, and $\lambda \sim
0.02-0.4$ for $\MBH=4\times10^7\Msun$ and $\cs\simgt 15\kms$.  For
classical Seyfert galaxies with strong bars, this suggests that the
masses of central BHs are likely to be less than $10^8\Msun$, which
appears to be consistent with the measured values from the relation
between the BH masses and the velocity dispersions of stellar bulges
(e.g., \citealt{wat08}) and reverberation mapping techniques (e.g.,
\citealt{gul09,den10}). Of course, this result may depend on many
factors such as the axis ratio and strength of the bar, presence of
self-gravity and magnetic fields, gas cooling and heating,
turbulence, etc., all of which would affect gas dynamic
significantly.  Extending the present work to include these physical
ingredients would be an important direction of future research.

\acknowledgments We gratefully acknowledge helpful discussions with
H.\ Ann, J.\ Goodman, M.\ G.\ Lee, E.\ Ostriker, D.\ Richstone, S.\
Tremaine, \& J.-H.\ Woo, especially F.\ Combes on the shapes of
nuclear spirals and bar torque.
We are also grateful to W.\ Maciejewski for stimulating comments
on the manuscript and to the referee for a thoughtful report.
This work was supported by the
National Research Foundation of Korea (NRF) grant funded by the
Korean government (MEST), No.\ 2010-0000712.

\appendix

\section{The Galaxy Model}\label{sec:appen}

In this Appendix we describe the gravitational potential of the model galaxy
that maintains the rotation of the (non-self-gravitating) gas disk.
The potential is comprised of four component: stellar disk, bulge, bar, and BH.
For the disk, we take a Kuzmin-Toomre model with surface density
\begin{equation}\label{eq:disk}
\sigma(R) = \frac{v_0^2}{2\pi G R_0} \left( 1 + \frac{R^2}{R_0^2}
\right)^{-3/2},
\end{equation}
where $v_0$ and $R_0$ are constants \citep{kuz56,too63}.
The corresponding gravitational potential at the disk midplane is
\begin{equation}\label{eq:dpot}
\Phi_{\rm disk} = - \frac{v_0^2R_0}{\sqrt{R^2 + R_0^2}},
\end{equation}
\citep{bin08}.
We take $v_0 = 260\kms$ and $R_0=14.1\kpc$ in our simulations.\footnote{
Our choice of the disk model is slightly different from \citet{pin95}
in that they took $v_0=200\kms$ and a surface density profile that is
singular at $R=0$ (see their Eq.\ (1)). Our adopted parameters
nevertheless give the rotational curves similar to those shown
in their Figure 1.}
The total disk mass is $M_{\rm disk}=v_0^2 R_0/G=2.2\times 10^{11}\Msun$.

For the bulge, we use a modified Hubble profile with volume density
\begin{equation}\label{eq:bulge}
\rho(R) = \rhobul \left(1 + \frac{R^2}{R_b^2}\right)^{-3/2},
\end{equation}
and the potential
\begin{equation}
\Phi_{\rm bul} = - \frac{4\pi G \rhobul R_b^3}{R}
\ln\left(\frac{R}{R_b} + \sqrt{1 + \frac{R^2}{R_b^2}}\right),
\end{equation}
where $\rhobul$ and $R_b$ are the central density and the characteristic
size of the bulge, respectively.  We take $\rhobul=2.4\times 10^{10}\dunit$
and $R_b=0.33\kpc$, with the corresponding bulge mass of
$M_{\rm bul} = 2.8\times 10^{10}\Msun$ within $R=6\kpc$.

The bar is modeled by a \citet{fer87} ellipsoid with volume density
\begin{eqnarray}\label{eq:bar}
\rho =
\left\{ \begin{array}{ll}
\rhobar \left( 1-g^2 \right)^n & ~~~{\rm for}~ g<1, \\
~~~~~~~0 & ~~~{\rm elsewhere},
\end{array} \right.
\end{eqnarray}
where $\rhobar$ is the bar central density,
$g^2= y^2/a^2 +(x^2 +z^2)/b^2$ with $(x, y, z)$ being
the Cartesian coordinates, and $a$ and $b$ are the semimajor and
semiminor axes of the bar, respectively.
The exponent $n$ measures the central concentration of the bar density
distribution.  In this work,
we fix the bar parameters to
$n=1$, $a=5\kpc$, $b=2\kpc$, and $\rhobar= 4.5\times 10^8\dunit$
for all models.  The total mass of the bar is then
$M_{\rm bar} = 2^{2n+3}\pi a b^2\rhobar \Gamma(n+1)\Gamma(n+2)/\Gamma(2n+4)
=1.5\times 10^{10}\Msun$ and the bar quadrupole moment is
$Q_m = M_{\rm bar} a^2 [ 1 -(b/a)^2]/(5+2n) = 4.5\times 10^{10}\Msun\kpc^2$,
corresponding to a strong bar.
When $n=1$, the bar potential is given explicitly as
\begin{eqnarray}\label{eq:barp}
\Phi_{\rm bar}(x,y,z) = - \frac{\pi G ab^2 \rho_{\rm bar}}{2}
[ ~W_{000} + x^2(x^2W_{200} + 2y^2W_{110}-2W_{100}) &\nonumber \\
           + y^2(y^2W_{020} + 2z^2W_{011}-2W_{010}) &\nonumber \\
           + z^2(z^2W_{002} + 2x^2W_{101}-2W_{001}) &],
\end{eqnarray}
where the coefficients $W_{ijk}$'s are defined in \citet{pfe84}.
As the bar pattern speed, we choose $\Omb=33\freq$ which places
the corotation resonance at $\RCR=6\kpc$.

Finally, for a central BH with mass $\MBH$, we use a Plummer potential
\begin{equation}
\Phi_{\rm BH} = - \frac{G\MBH}{(R^2 + R_s^2)^{1/2}},
\end{equation}
with the softening radius $R_s=1$ pc.  We take $\MBH=0$,
$4\times10^7\Msun$, and $4\times10^8\Msun$ to study the effects of
the central mass concentration on the bar substructures. With $\MBH
\ll M_{\rm disk}$, $M_{\rm bul}$, $M_{\rm bar}$, the BH affects the
rotation curve only in the very central regions (with $R \simlt
0.5\kpc$).
Figures \ref{fig:rotcurve} and \ref{fig:angfreq} plot the
rotational velocity for models with $\MBH=4\times 10^8\Msun$ and the
angular frequency curves for all models, respectively.

\end{document}